\documentclass[prd,aps,a4paper,superscriptaddress,twocolumn,footinbib,showpacs]{revtex4}

\usepackage{latexsym}

\newif\ifAMS
\IfFileExists{amssymb.sty}
  {\AMStrue\usepackage{amssymb}}
  {\usepackage{latexsym}}

\newif\ifST
\IfFileExists{stmaryrd.sty}
  {\STtrue\usepackage{stmaryrd}}
  {}

\newif\ifEU
\IfFileExists{euscript.sty}
  {\EUtrue\usepackage[mathcal]{euscript}}
  {}

\usepackage{graphicx}
\usepackage{amsmath}
\usepackage{color}
\usepackage{dcolumn}
\usepackage{bm}
\usepackage{slashed}
\allowdisplaybreaks
\begin{document}
\title{A waveform model for eccentric binary black hole based on effective-one-body-numerical-relativity (EOBNR) formalism}

\author{Zhoujian Cao}
\email{zjcao@amt.ac.cn} \affiliation{Department of Astronomy, Beijing Normal University, Beijing 100875, China}
\affiliation{Institute of Applied
Mathematics, Academy of Mathematics and Systems Science,
Chinese Academy of Sciences, Beijing 100190, China}
\author{Wen-Biao Han}
\affiliation{Shanghai Astronomical Observatory, 80 Nandan Road,
Shanghai, 200030, People's Republic of China}
\affiliation{School of Astronomy and Space Science, University of Chinese Academy of Sciences,  Beijing 100049, China}

\begin{abstract}
Binary black hole systems are among the most important sources for gravitational wave detection. And also they
are good objects for theoretical research for general relativity. Gravitational waveform template is important to data analysis. Effective-one-body-numerical-relativity (EOBNR) model has played an essential role in the LIGO data analysis. For future space-based gravitational wave detection, many binary systems will admit somewhat orbit eccentricity. At the same time the eccentric binary is also an interesting topic for theoretical study in general relativity. In this paper we construct the first eccentric binary waveform model based on effective-one-body-numerical-relativity framework. Our basic assumption in the model construction is that the involved eccentricity is small. We have compared our eccentric EOBNR model to the circular one used in LIGO data analysis. We have also tested our eccentric EOBNR model against to another recently proposed eccentric binary waveform model; against to numerical relativity simulation results; and against to perturbation approximation results for extreme mass ratio binary systems. Compared to numerical relativity simulations with eccentricity as large as about 0.2, the overlap factor for our eccentric EOBNR model is better than 0.98 for all tested cases including spinless binary and spinning binary; equal mass binary and unequal mass binary. Hopefully our eccentric model can be the start point to develop a faithful template for future space-based gravitational wave detectors.
\end{abstract}


\maketitle


\section{Introduction}
The direct detection of gravitational waves (GW) has been announced recently by LIGO \cite{Abbott:2016blz,Abbott:2016nmj,PhysRevLett.118.221101} which opens the brand new window to our universe--gravitational wave astronomy. The success of LIGO is based on both the tremendous development of experiment technology and the improvement of theoretical research in the past decades. Matched filtering data analysis technique is very important to gravitational wave detection. As GW150914, GW151226 and GW170104 have witnessed, the matched filtering technique has improved the data quality and/or even make the noisy data detectable. Regarding to GW150914 and GW170104, we are some lucky. The signal is so strong that the matched filtering data analysis technology is not necessary to catch the signal, although the matched filtering data analysis can improve the signal to noise ratio (SNR) and confidence level strongly. Regarding to GW151226, the signal is much weaker than that of GW150914 and GW170104. Without the matched filtering data analysis technology, the signal is completely invisible. In contrast, the matched filtering data analysis technology digs out the signal from the strong noise with SNR 13 and confidence level 5$\sigma$. GW151226 is a good example showing that the detection of GW is the result of combination of experiment achievement and theoretical research progress \cite{cao2016gravitational}.

In order to make the matched filtering technique work, the gravitational waveform template is essential \cite{cao2016gravitational}. And the template strongly depends on the specific theoretical model of GW source. Currently there are two theoretical models which are ready for gravitational wave data analysis. They are effective-one-body-numerical-relativity (EOBNR) model \cite{PhysRevD.89.061502} and IMRPhenom model \cite{PhysRevLett.116.241102}. For example, all of GW150914, GW151226 and GW170104 depend on these two models strongly.

EOBNR model \cite{PhysRevD.76.104049} is a combination of effective-one-body theory of post-Newtonian approximation and numerical relativity. About the template bank of the binary black hole gravitational waveform, the related parameters are divided into intrinsic ones and extrinsic ones. The EOBNR model need only concern the intrinsic parameters including the total mass of the binary black hole $M$, the mass ratio $q\equiv\frac{m_1}{m_2}\geq1$, the spins of the two black holes $\vec{S}_{1,2}$ and the eccentricity of the orbit $e$. While the extrinsic parameters, including luminosity distance $D$, source location $(\theta,\phi)$, the configuration of the orbit respect to the sight direction $(\iota,\beta)$, the reaching time of the signal $t_0$, the initial phase $\phi_0$ and the polarization angle respect to the detector $\psi$, can be straightforwardly involved when we construct the template bank from EOBNR model.

To the quasi-circular $e=0$ and non-precession ($\vec{S}_{1,2}$ perpendicular to the orbit plane) binary black hole systems with mass ratio $q\in(1,\frac{20}{1})$, the EOBNR model has done a quite good job \cite{PhysRevD.89.061502}. (Note that the author in \cite{PhysRevD.93.064041} mentioned the EOBNR model is valid for mass ratio range $1<q<100$ and spin range $-1<\chi<0.99$. But as pointed out by \cite{PhysRevD.93.044006,PhysRevD.93.044007,chu2016accuracy,PhysRevD.93.104050}, current EOBNR model can at most be calibrated to numerical relativity only for the range $1<q<20$ and $-0.85<\chi<0.85$.) Regarding to precession binaries, a primary development of EOBNR model is available \cite{pan2014inspiral}. Very recently, we have done an initial investigation to extend the EOBNR model to include gravitational wave memory in reference \cite{Cao16}. As to the mass ratio, there is no essential difficult to extend the EOBNR model to cover larger parameter range. But current simulation power of numerical relativity limits such development \cite{PhysRevLett.106.041101}. In principle, if only relevant numerical relativity results are available the EOBNR model can be calibrated to involved mass ratio. Regarding to eccentricity, the situation is different. Till now, the EOBNR model only works for $e=0$ case. Reference \cite{PhysRevD.86.124012} touched this problem, but the authors only considered energy flux while left the relevant gravitational wave form alone. Although the EOBNR model admits kinds of limitations as described above, it provides a framework which is possible to extend the EOBNR model to treat these limitations. Recently the authors in \cite{julie2017two} have extended EOB framework to scalar-tensor theory. Hopefully ones can treat gravitational waveform template for different gravitational theories \cite{PhysRevD.94.084002} within one uniform framework, EOBNR model, in the future.

Due to the circularization effect of gravitational radiation \cite{PhysRev.136.B1224}, ones may expect that the binary black hole systems are always near circular when they enter the LIGO frequency band. But recent investigations show it is not absolutely true. The study of the galactic cluster M22 indicate that about 20\% of binary black hole (BBH) mergers in globular clusters will have eccentricities larger than 0.1 when they first enter advanced LIGO band at 10Hz \cite{strader2012two}, and that $\sim$10\% may have eccentricities $e\sim1$ \cite{antonini2016black}. Furthermore, a fraction of galactic field binaries may retain significant eccentricity prior to the merger event \cite{samsing2014formation}. BBHs formed in the vicinity of supermassive black holes (BH) may also merge with significant residual eccentricities \cite{vanlandingham2016role}. For space-based detectors such as eLISA \cite{amaro2012low}, LISA \cite{bender1998lisa,audley2017laser}, Taiji \cite{gong2011scientific} and Tianqin \cite{luo2016tianqin}, the orbit of the involved binary black hole systems may be highly eccentric due to recent perturbations by other orbiting objects \cite{hils1995gradual,PhysRevD.50.6297}. Recently there are many authors care about the binary black hole systems with eccentric orbit regarding to gravitational wave detection \cite{PhysRevD.80.084001,PhysRevD.90.084016,PhysRevD.92.044034,PhysRevD.95.024038}.

Assuming low eccentricity, the authors of \cite{PhysRevD.80.084001} extended low order PN waveform model in frequency domain to including eccentricity. They called the corresponding model post-circular (PC) model. Later, the authors of \cite{PhysRevD.90.084016} improved the PC model to EPC (enhanced post-circular) model which recovers the TaylorF2 model when the eccentricity vanishes. The EPC model is a phenomenologically extending of PC model. Its overall PN order is 3.5. Some numerical relativity simulations have been paid in the past to the eccentric binary black hole systems \cite{PhysRevD.78.064069,PhysRevD.82.024033,PhysRevD.88.064051}. Along with the numerical relativity results, the x-model was proposed in \cite{PhysRevD.82.024033}. The x-model is a low order post-Newtonian (PN) model. Recently, this model was improved to include inspiral, merger and ringdown phases, and higher PN order terms for vanishing eccentricity part are included. This model was called ax-model by the authors of \cite{PhysRevD.95.024038}. All these models are valid for any mass ratio.

Regarding to large mass ratio binary black hole systems, ones may look the binary system as a perturbation of the big black hole. Then the gravitational wave problem is decomposed into the trajectory problem and the related waveform problem. In \cite{han2014gravitational}, one of us investigated the eccentric binary using the Teukolsky equation to treat the waveform problem and combining the conserved EOB dynamics with numerical energy flux to treat the trajectory \cite{han2011constructing}. In \cite{han2011constructing,han2014gravitational} the Teukolsky equation is solved numerically. Ones can also solve it through some analytical method \cite{sasaki2003analytic} or post-Newtonian approximation \cite{PhysRevD.93.064058}. In \cite{PhysRevD.73.024027}, the authors used geodesic equation to treat the eccentric orbit of a large mass ratio binary and used the Teukolsky equation to treat the waveform problem. Interestingly, people have used the method of geodesic equation and Teukolsky equation to find that the eccentricity may increase \cite{PhysRevD.47.5376,PhysRevD.58.064012} instead of always decay found through post-Newtonian approximation \cite{PhysRev.136.B1224}. And more, people used the method of geodesic equation and Teukolsky equation to find the interesting transient resonance phenomena \cite{PhysRevLett.109.071102,PhysRevD.94.124042}. When a binary system passes through a transient resonance, the radial frequency and polar frequency become commensurate, and the orbital parameters will show a jump behavior. To our knowledge, the post-Newtonian approximation method can not yet give out the eccentricity increasing and the transient resonance results. Of course it is possible that the available post-Newtonian result is not accurate enough to get these two interesting phenomena. But it is also deserved to ask whether these two phenomena imply that the perturbation method breaks down. Ideally ones may use numerical relativity simulation to check this problem. But unfortunately, current numerical relativity techniques are far away to investigate this problem due to the huge computational cost for large mass ratio binary systems \cite{PhysRevLett.106.041101} (but see \cite{lewis2016fundamental}). Hopefully, effective-one-body-numerical-relativity (EOBNR) model may be used to check this problem. This is because on the side of almost equal mass cases, EOBNR framework can be and has been calibrated against numerical relativity; on the side of extreme mass ratio cases, EOB framework can also be and has also been used to describe the dynamics and the gravitational waveform \cite{PhysRevLett.104.091102}. So we can expect EOBNR framework may play a bridge role to connect numerical relativity result with large mass ratio problem. In order to realize this kind of investigation, we need a EOBNR model being valid for eccentric binary systems, which is absent now. In current paper, we will go a little step to construct an EOBNR model for eccentric binary systems.

This paper is organized as follows. In the next section we will describe the extended EOBNR model including eccentricity. We call our model SEOBNRE (Spinning Effective-One-Body-Numerical-Relativity model for Eccentric binary). This model includes three essential parts which will be explained in detail respectively in the subsections of next section. The involved detail calculations and long equations are postponed to the Appendix. Then in Sec.~\ref{sec::tests} we check and test our SEOBNRE model against to quasi-circular EOBNR model; against to existing eccentric waveform model--ax model; against to numerical relativity simulation results and against to Teukolsky equation based waveform model for extreme mass ratio binary systems. Finally we give a summary and discussion in Sec.~\ref{sec::discussion}. Throughout this paper we will use units $c=G=1$. Regarding to the mass of the binary we always assume $m_1\geq m_2$.

\section{waveform model for eccentric binary based on EOBNR}
Effective one body technique is a standard trick to treat the two body problem in the central force situation of classical mechanics, especially for Newtonian gravity theory \cite{goldstein2002classical}. In \cite{buonanno1999effective}, Buonanno and Damour introduced the seminal idea of effective-one-body approach for general relativistic two body problem. The effective-one-body approach needs many inputs from post-Newtonian approximation, but it is more powerful than post-Newtonian approximation. Not like post-Newtonian approximation which will diverge before the late inspiral stage of the binary evolution, the effective-one-body approach works till the binary merger. And more it is convenient for the effective-one-body approach to adopt the result of perturbation method \cite{PhysRevLett.104.091102}. At the same time, we can also combine the results of the effective-one-body approach and numerical relativity. As firstly done by Pan and his coworkers in \cite{PhysRevD.76.104049}, such combination gives effective-one-body numerical relativity (EOBNR) model. Currently the most advanced EOBNR model is the SEOBNR which includes version 1 \cite{PhysRevD.86.024011}, version 2 \cite{PhysRevD.89.061502} and version 3 \cite{PhysRevX.6.041014,PhysRevD.95.024010,PhysRevLett.118.221101}. SEOBNR is valid only for quasi-circular orbit and black hole spin perpendicular to the orbital plane which means the precession is not presented. In this paper, we will extend SEOBNR model to treat eccentric orbit.

The EOB approach includes three building parts: (1) a description of the conservative part of the dynamics of two compact bodies which is represented by a Hamiltonian; (2) an expression for the radiation-reaction force which is added to the conservative Hamiltonian equations of motion; and (3) a description of the asymptotic gravitational waveform emitted by the binary system. The part 1 is independent of the character of the involved orbit. In another word, the part 1 is valid no matter the orbit is circular or eccentric. In current paper, we adopt the result from SEOBNRv1 model which will be summarized in the following. Regarding to parts 2 and 3, current SEOBNRv1 model is not valid to eccentric orbit. We will extend these two parts in current work. For convenience, we will refer our model SEOBNRE where the last letter E represents eccentricity.

\subsection{Conservative part for SEOBNRE model}
The conservative part for SEOBNRE model is the same to that of SEOBNRv1 \cite{PhysRevD.86.024011}. But the related equations are distributed in different papers. For reference convenience, we give a summary here.

The basic idea of EOB approach is reducing the conservative dynamics of the two body problem in general relativity to a geodesic motion (more precisely Mathisson-Papapetrou-Dixon equation \cite{PhysRevD.80.104025}) on the top of a reduced spacetime which corresponds to the reduced one body. Roughly the reduced spacetime is a deformed Kerr black hole with metric form \cite{PhysRevD.81.084024}
\begin{align}
ds^2&=g_{tt}dt^2+g_{rr}dr^2+g_{\theta\theta}d\theta^2+g_{\phi\phi}d\phi^2+2g_{t\phi}dtd\phi,\\
g^{tt}&=-\frac{\Lambda_t}{\Delta_t\Sigma},\\
g^{rr}&=\frac{\Delta_r}{\Sigma},\\
g^{\theta\theta}&=\frac{1}{\Sigma},\\
g^{\phi\phi}&=\frac{1}{\Lambda_t}(-\frac{\tilde{\omega}_{fd}^2}{\Delta_t\Sigma}+\frac{\Sigma}{\sin^2\theta}),\\
g^{t\phi}&=-\frac{\tilde{\omega}_{fd}}{\Delta_t\Sigma}
\end{align}
where
\begin{align}
\Sigma&=r^2+a^2\cos^2\theta,\\
\Delta_t&=r^2(A(u)+\frac{a^2u^2}{M^2}),\\
\Delta_r&=\frac{\Delta_t}{D(u)},\\
\Lambda_t&=\bar{\omega}^4-a^2\Delta_t\sin^2\theta,\\
\bar{\omega}&\equiv\sqrt{r^2+a^2},\\
\tilde{\omega}_{fd}&=2Mar+\frac{Ma\eta}{r}(\omega^{fd}_1M^2+\omega^{fd}_2a^2).
\end{align}
We still call the coordinate $(t,r,\theta,\phi)$ used here Boyer-Lindquist coordinate. Following SEOBNRv1 we set $\omega^{fd}_1=0$ and $\omega^{fd}_2=0$. Here $M$ and $a$ are respectively the mass and the Kerr spin parameter of the deformed Kerr black hole
\begin{align}
M&\equiv m_1+m_2,\\
M\vec{a}&\equiv\vec{\sigma} \equiv \vec{a}_1m_1+\vec{a}_2m_2
\end{align}
and we have used notation $u\equiv\frac{M}{r}$ and
\begin{align}
A(u)&\equiv1-2u+2\eta u^3+\eta(\frac{94}{3}-\frac{41}{32}\pi^2)u^4,\\
D(u)&\equiv1/\{1+\log[1+6\eta u^2+2(26-3\eta)\eta u^3]\},
\end{align}
where $\eta\equiv\frac{m_1m_2}{(m_1+m_2)^2}$ is the symmetric mass ratio of the binary with components mass $m_1$, $m_2$ and Kerr parameter $\vec{a}_1$ and $\vec{a}_2$.

Corresponding to the geodesic motion, or to say the Mathisson-Papapetrou-Dixon equations, the Hamiltonian can be written as \cite{PhysRevD.84.104027,PhysRevD.86.024011}
\begin{align}
H&=M\sqrt{1+2\eta(\frac{H_{eff}}{M\eta}-1)},\label{SEOBNREHtotal}\\
H_{eff}&=H_{NS}+H_S+H_{SC}.\label{SEOBNREHterms}
\end{align}
The detail expressions for the quantities $H_{NS}$, $H_{S}$ and $H_{SC}$ involved in the above Hamiltonian are listed in the Appendix~\ref{App:ham}.

Based on the above given Hamiltonian, we have then the equation of motion respect to the conservative part as
\begin{align}
\dot{\vec{r}}&=\frac{\partial H}{\partial\vec{\tilde{p}}},\label{dynr}\\
\dot{\vec{\tilde{p}}}&=-\frac{\partial H}{\partial\vec{r}}.\label{dynp}
\end{align}
\subsection{Gravitational waveform part for SEOBNRE model}
In the EOBNR framework, the gravitational wave form is described by spin-weighted $-2$ spherical harmonic modes. This kind of modes is also extensively used in numerical relativity \cite{bai11}. In SEOBNRv1, the modes $\ell\in\{2,3,4,5,6,7,8\},m\in[-\ell,\ell]$ are available. Note that only the positive $m$ modes are considered while the negative $m$ modes are produced through relation $h_{\ell m}=(-1)^\ell h_{\ell,-m}^*$ \cite{PhysRevD.84.124052}. Here $*$ means the complex conjugate.

In this work, we only consider $(\ell,m)=(2,2)$ mode although other modes can be extended straightforward. Our basic idea is decomposing the waveform into quasi-circular part and eccentric part. The strategy is following \cite{PhysRevD.95.024038}. We treat the eccentric part as perturbation by assuming that the eccentricity is small. Regarding to the quasi-circular part, we borrow the ones from SEOBNRv1 exactly. For convenience, we firstly review this part. Within EOBNR framework, the waveform is divided into two segments. One is after merger which is described with quasi-normal modes of some Kerr black hole. The other is inspiral-plunge stage which is described in the factorized form as \cite{PhysRevD.84.124052}
\begin{align}
h_{\ell m}^{(C)}&=h_{\ell m}^{(N,\epsilon)}\hat{S}_{eff}^{(\epsilon)}T_{\ell m}e^{i\delta_{\ell m}}(\rho_{\ell m})^\ell N_{\ell m},\label{CEOB}\\
h_{\ell m}^{(N,\epsilon)}&=\frac{M\eta}{R}n_{\ell m}^{(\epsilon)}c_{\ell+\epsilon}V_\Phi^\ell Y^{\ell-\epsilon,-m}(\frac{\pi}{2},\Phi),
\end{align}
where $R$ is the distance to the source; $\Phi$ is the orbital phase; $Y^{\ell m}(\Theta,\Phi)$ are the scalar spherical harmonics. Particularly for $(2,2)$ mode, we have \cite{PhysRevD.83.064003,PhysRevD.84.124052,PhysRevD.86.024011}
\begin{align}
\epsilon&=0,\\
V_\Phi^2&=v_\Phi^2,\\
\vec{v}_\Phi&=\vec{v}_p-\vec{n}(\vec{v}_p\cdot\vec{n}),\vec{v}_p\equiv\dot{\vec{r}}=\frac{\partial H}{\partial\vec{\tilde{p}}},\\
n^{(0)}_{\ell m}&=(im)^\ell\frac{8\pi}{(2\ell+1)!!}\sqrt{\frac{(\ell+1)(\ell+2)}{\ell(\ell-1)}},\\
c_{\ell}&=(\frac{m_2}{m_1+m_2})^{\ell-1}+(-1)^\ell(\frac{m_1}{m_1+m_2})^{\ell-1},\\
\hat{S}_{eff}^{(0)}&=\frac{H_{eff}}{M\eta},\\
T_{\ell m}&=\frac{\Gamma(\ell+1-2im\Omega H)}{\Gamma(\ell+1)}e^{\pi m\Omega H+2im\Omega H\ln(2m\Omega r_0)},\\
\Omega&\equiv v_\Phi^3, r_0\equiv\frac{2(m_1+m_2)}{\sqrt{e}},\\
N_{\ell m}&=[1+\frac{\tilde{p}_r^2}{(r\Omega)^2}(a_1^{h_{\ell m}}+\frac{a_2^{h_{\ell m}}}{r}+\frac{a_3^{h_{\ell m}}+a_{3S}^{h_{\ell m}}}{r^{3/2}}\nonumber\\
&+\frac{a_4^{h_{\ell m}}}{r^2}+\frac{a_5^{h_{\ell m}}}{r^{5/2}})]\exp[i(\frac{\tilde{p}_r}{r\Omega}b_1^{h_{\ell m}}\nonumber\\
&+\frac{\tilde{p}_r^3}{r\Omega}(b_2^{h_{\ell m}}
+\frac{b_3^{h_{\ell m}}}{r^{1/2}}+\frac{b_4^{h_{\ell m}}}{r}))],\\
\delta_{22}&=\frac{7}{3}\bar{v}^3+(\frac{428}{105}\pi-\frac{4}{3}a)\bar{v}^6+
(\frac{1712}{315}\pi^2-\frac{2203}{81})\bar{v}^9\nonumber\\
&-24\eta v_\Phi^5+\frac{20}{63}av_\Phi^8,\\
\rho_{22}&=1+(\frac{55}{84}\eta-\frac{43}{42})v_\Phi^2\nonumber\\
&-\frac{2}{3}[\chi_S(1-\eta)+\chi_A\sqrt{1-4\eta}]v_\Phi^3\nonumber\\
&+(\frac{a^2}{2}-\frac{20555}{10584}-\frac{33025}{21168}\eta+\frac{19583}{42336}\eta^2)v_\Phi^4-\frac{34}{21}av_\Phi^5\nonumber\\
&+[\frac{1556919113}{122245200}+\frac{89}{252}a^2-\frac{48993925}{9779616}\eta-\frac{6292061}{3259872}\eta^2\nonumber\\
&+\frac{10620745}{39118464}\eta^3+\frac{41}{192}\eta\pi^2-\frac{428}{105}\text{eulerlog}_2(v_\Phi^2)]v_\Phi^6\nonumber\\
&+(\frac{18733}{15876}a+\frac{1}{3}a^3)v_\Phi^7\nonumber\\
&+[\frac{18353}{21168}a^2-\frac{1}{8}a^4-(5.6+117.6\eta)\eta\nonumber\\
&-\frac{387216563023}{160190110080}+\frac{9202}{2205}\text{eulerlog}_2(v_\Phi^2)]v_\Phi^8\nonumber\\
&-[\frac{16094530514677}{533967033600}-\frac{439877}{55566}\text{eulerlog}_2(v_\Phi^2)]v_\Phi^{10},\\
\chi_S&=\frac{a_1/m_1+a_2/m_2}{2},\chi_A=\frac{a_1/m_1-a_2/m_2}{2}
\end{align}
where we have defined $\text{eulerlog}_m(v_\Phi^2)\equiv\gamma_E+\ln(2mv_\Phi)$ with $\gamma_E\approx0.5772156649015328606065120900824024$ being the Euler constant. In the equation of $N_{\ell m}$, the parameters $a_{1}^{h_{\ell m}}$, $a_{2}^{h_{\ell m}}$, $a_{3}^{h_{\ell m}}$, $b_{1}^{h_{\ell m}}$, $b_{2}^{h_{\ell m}}$ are functions of $\eta$, and parameters $a_{3S}^{h_{\ell m}}$, $a_{4}^{h_{\ell m}}$, $a_{5}^{h_{\ell m}}$, $b_3^{h_{\ell m}}$ and $b_4^{h_{\ell m}}$ are functions of $a$ and $\eta$. Following SEOBNRv1, we construct data tables for $a_{i}^{h_{\ell m}}$, $a_{3S}^{h_{\ell m}}$, $b_{1}^{h_{\ell m}}$ and $b_{2}^{h_{\ell m}}$ based on the numerical relativity results of some specific cases for $a$ and $\eta$. Then we interpolate to get the wanted values for the $a$ and $\eta$ in question. Then we solve the conditions (21)-(25) of \cite{PhysRevD.86.024011} for $b_3^{h_{\ell m}}$ and $b_4^{h_{\ell m}}$.

For the eccentric part, the post-Newtonian (PN) result is valid till second PN order \cite{PhysRevD.54.4813}
\begin{align}
&h^{ij}=2\eta(Q^{ij}+P^{\frac{1}{2}}Q^{ij}+PQ^{ij}
+P^{\frac{3}{2}}Q^{ij}\nonumber\\
&+P^{\frac{3}{2}}Q_{\text{tail}}^{ij}),\label{ewaveb}\\
&Q^{ij}=2(v_p^iv_p^j-\frac{n^in^j}{r}),\\
&P^{\frac{1}{2}}Q^{ij}=(m_1-m_2)[\frac{3N_n}{r}(n^iv_p^j+v_p^in^j-\dot{r}n^in^j)\nonumber\\
&+N_v(\frac{n^in^j}{r}-2v_p^iv_p^j)],\\
&PQ^{ij}=\frac{1}{3}\{(1-3\eta)[\frac{N_n^2}{r}((3v_p^2-15\dot{r}^2+\frac{7}{r})n^in^j\nonumber\\
&+15\dot{r}(n^iv_p^j+v_p^in^j)-14v_p^iv_p^j)
+\frac{N_nN_v}{r}(12\dot{r}n^in^j\nonumber\\
&-16(n^iv_p^j+v_p^in^j))+N_v^2(6v_p^iv_p^j-\frac{2}{r}n^in^j)]\nonumber\\
&+[3(1-3\eta)v_p^2-2\frac{(2-3\eta)}{r}]v_p^iv_p^j\nonumber\\
&+\frac{2}{r}\dot{r}(5+3\eta)(n^iv_p^j+v_p^in^j)\nonumber\\
&+[3(1-3\eta)\dot{r}^2-(10+3\eta)v_p^2+\frac{29}{r}]\frac{n^in^j}{r}\},\\
&P^{\frac{3}{2}}Q^{ij}=(m_1-m_2)(1-2\eta)\{\frac{N_n^3}{r}[\frac{5}{4}
(3v_p^2-7\dot{r}^2+\frac{6}{r})\dot{r}n^in^j\nonumber\\
&-\frac{17}{2}\dot{r}v_p^iv_p^j-(21v_p^2-105\dot{r}^2
+\frac{44}{r})\frac{n^iv_p^j+v_p^in^j}{12}]\nonumber\\
&+\frac{1}{4}\frac{N_n^2N_v}{r}[58v_p^iv_p^j+(45\dot{r}^2-9v_p^2-\frac{28}{r})n^in^j\nonumber\\
&-54\dot{r}(n^iv_p^j+v_p^in^j)]
+\frac{3}{2}\frac{N_nN_v^2}{r}(5(n^iv_p^j+v_p^in^j)-3\dot{r}n^in^j)\nonumber\\
&+\frac{1}{2}N_v^3(\frac{n^in^j}{r}-4v_p^iv_p^j)\}+\frac{\delta m}{12}\frac{N_n}{r}\{(n^iv_p^j\nonumber\\
&+v_p^in^j)[\dot{r}^2(63+54\eta)-\frac{128-36\eta}{r}
+v_p^2(33-18\eta)]\nonumber\\
&+n^in^j\dot{r}[\dot{r}^2(15-90\eta)-v_p^2(63-54\eta)+\frac{242-24\eta}{r}]\nonumber\\
&-\dot{r}v_p^iv_p^j(186+24\eta)\}+(m_1-m_2)N_v\{\frac{1}{2}v_p^iv_p^j[\frac{3-8\eta}{r}\nonumber\\
&-2v_p^2(1-5\eta)]-\frac{n^iv_p^j+v_p^in^j}{2r}\dot{r}(7+4\eta)\nonumber\\
&-\frac{n^in^j}{r}[\frac{3}{4}(1-2\eta)\dot{r}^2
+\frac{1}{3}\frac{26-3\eta}{r}-\frac{1}{4}(7-2\eta)v_p^2]\},\\
&P^{\frac{3}{2}}Q_{\text{tail}}^{ij}=4v_p^5[\pi(\lambda^i\lambda^j-n^in^j)
+6\ln{v_p}(\lambda^in^j+n^i\lambda^j)].
\end{align}
In the above equations we have used the following notations. $\hat{N}=(\sin\theta\cos\phi,\sin\theta\sin\phi,\cos\phi)$ is the radial direction to the observer. $\hat{p}=(\cos\theta\cos\phi,\cos\theta\sin\phi,-\sin\theta)$ lies along the line of nodes. $\hat{q}=\hat{N}\times\hat{p}$ and more notations include \cite{PhysRevD.54.4813,PhysRevD.52.821}
\begin{align}
N_n&=\hat{N}\cdot \vec{n},\\
N_v&=\hat{N}\cdot\vec{v_p},\\
\vec{\lambda}&=\frac{\vec{v_p}-(\vec{v_p}\cdot\vec{n})\vec{n}}
{|\vec{v_p}-(\vec{v_p}\cdot\vec{n})\vec{n}|}.
\end{align}
We define the spin-weighted spherical harmonic modes as
\begin{align}
\epsilon^+_{ij}&=\frac{1}{2}(\hat{p}_i\hat{p}_j-\hat{q}_i\hat{q}_j),\\
\epsilon^\times_{ij}&=\frac{1}{2}(\hat{p}_i\hat{q}_j+\hat{q}_i\hat{p}_j).\\
h_+&=\epsilon^+_{ij}h^{ij},\\
h_\times&=\epsilon^\times_{ij}h^{ij}.\\
h&=h_+-ih_\times,\\
h_{\ell,m}&=\int h {}^{-2}Y^*_{\ell,m}d\Omega.\label{ewavee}
\end{align}
Based on above results we express the (2,2) mode as
\begin{align}
h_{22}&=2\eta[\Theta_{ij}(Q^{ij}+P_0Q^{ij}+P^{\frac{3}{2}}Q^{ij}_{\text{tail}})\nonumber\\
&+P_n\Theta_{ij}(P_n^{\frac{1}{2}}Q^{ij}+P^{\frac{3}{2}}_nQ^{ij})\nonumber\\
&+P_v\Theta_{ij}(P_v^{\frac{1}{2}}Q^{ij}+P^{\frac{3}{2}}_vQ^{ij})\nonumber\\
&+P_{nn}\Theta_{ij}P_{nn}Q^{ij}\nonumber\\
&+P_{nv}\Theta_{ij}P_{nv}Q^{ij}\nonumber\\
&+P_{vv}\Theta_{ij}P_{vv}Q^{ij}\nonumber\\
&+P_{nnn}\Theta_{ij}P^{\frac{3}{2}}_{nnn}Q^{ij}\nonumber\\
&+P_{nnv}\Theta_{ij}P^{\frac{3}{2}}_{nnv}Q^{ij}\nonumber\\
&+P_{nvv}\Theta_{ij}P^{\frac{3}{2}}_{nvv}Q^{ij}\nonumber\\
&+P_{vvv}\Theta_{ij}P^{\frac{3}{2}}_{vvv}Q^{ij}].\label{h22will}
\end{align}
The involved notations such $\Theta_{ij}$ and $P_n\Theta_{ij}$ are explained one by one in the Appendix~\ref{App:waveform}.

We assume the $h_{22}$ in the equation (\ref{h22will}) includes quasi-circular part corresponding to $h_{22}|_{\dot{r}=0}$ and the left eccentric part. It is straightforward to check that $h_{22}|_{\dot{r}=0}$ is consistent to the Eq.~(9.3) of \cite{Blanchet16}. So we define the eccentric correction as
\begin{align}
h_{22}^{(PNE)}=h_{22}-h_{22}|_{\dot{r}=0},\label{h22willelip}
\end{align}
where $h_{22}$ means the one given in the equation (\ref{h22will}). In summary the inspiral-plunge waveform for SEOBNRE is
\begin{align}
h_{22}^{insp-plun}=h_{22}^{(C)}+h_{22}^{(PNE)}, \label{Ewave}
\end{align}
where $h_{22}^{(C)}$ is given in Eq.~(\ref{CEOB}).

\subsection{Radiation-reaction force for SEOBNRE model}
We have mentioned conservative part of the EOB dynamics in Eqs.~(\ref{dynr}) and (\ref{dynp}). But that is only partial part of the whole EOB dynamics. The left part is related to the radiation-reaction force. Assume the radiation-reaction force is $\vec{\mathcal{F}}$, then the whole EOB dynamics can be expressed as
\begin{align}
\dot{\vec{r}}&=\frac{\partial H}{\partial\vec{\tilde{p}}},\label{fdynr}\\
\dot{\vec{\tilde{p}}}&=-\frac{\partial H}{\partial\vec{r}}+\vec{\mathcal{F}}.\label{fdynp}
\end{align}
In SEOBNRv1 model, the radiation-reaction force $\vec{\mathcal{F}}$ is related to the energy flux of gravitational radiation $\frac{dE}{dt}$ through \cite{PhysRevD.86.024011}
\begin{align}
\vec{\mathcal{F}}&=\frac{1}{M\eta\omega_\Phi|\vec{r}\times\vec{\tilde{p}}|}\frac{dE}{dt}\vec{\tilde{p}},\\
\omega_\Phi&=\frac{|\vec{r}\times\dot{\vec{r}}|}{r^2}.
\end{align}
Here we need to note the sign of $\frac{dE}{dt}$. Since $E$ here means the energy of the binary system, $E$ decreases due to the gravitational radiation, $\frac{dE}{dt}<0$. So people call it dissipation sometimes. Corresponding to SEOBNRv1 code \cite{SEOBNRv1code}, since it treats quasi-circular cases without precession, $|\vec{r}\times\vec{\tilde{p}}|\approx \tilde{p}_\phi$ which reduces
\begin{align}
\vec{\mathcal{F}}&=\frac{1}{M\eta\omega_\Phi}\frac{dE}{dt}\frac{\vec{\tilde{p}}}{\tilde{p}_\phi}.
\end{align}

Regarding to the energy flux $\frac{dE}{dt}$, SEOBNR model relates it to the gravitational wave form through \cite{PhysRevD.83.064003}
\begin{align}
-\frac{dE}{dt}=\frac{1}{16\pi}\sum_\ell\sum_{m=-\ell}^{\ell}|\dot{h}_{\ell m}|^2.
\end{align}
And more the SEOBNR model assumes the dependence of $h_{\ell m}$ on time is a harmonic oscillation. So $\dot{h}_{\ell m}\approx m\Omega h_{\ell m}$ with $\Omega$ the orbital frequency of the binary. Then
\begin{align}
-\frac{dE}{dt}&=\frac{1}{16\pi}\sum_\ell\sum_{m=-\ell}^{\ell}(m\Omega)^2|h_{\ell m}|^2\\
&=\frac{1}{8\pi}\sum_\ell\sum_{m=1}^{\ell}(m\Omega)^2|h_{\ell m}|^2.\label{waveformflux}
\end{align}
Like EOBNR models, our SEOBNRE model is valid only for spin aligned binary black holes (spin is perpendicular to the orbital plane). In these systems, there is no precession will be involved. Most importantly, these systems admit a plane reflection symmetry respect to the orbital plane. Due to this symmetry of the corresponding spacetime and the symmetry of the spin-weighted spherical harmonic functions we have (Eqs.~(44)-(46) of \cite{PhysRevD.77.024027})
\begin{align}
h(t,\pi-\theta,\phi)&=h^*(t,\theta,\phi),\\
{}^{-2}Y_{\ell,-m}(\pi-\theta,\phi)&={}^{-2}Y^*_{\ell m}(\theta,\phi).
\end{align}
Here we have used $(\theta,\phi)$ to represent the spherical coordinate respect to the gravitational wave source. Consequently we have $h_{\ell m}=(-1)^\ell h_{\ell,-m}^*$ \cite{PhysRevD.84.124052}. In the second equality of the above energy flux equation, SEOBNR model has taken this relation into consideration and has neglected the `memory' modes $h_{\ell 0}$ \cite{Cao16,PhysRevD.95.084048}. We also note that some authors use the relation $h_{\ell m}=(-1)^\ell h_{\ell,-m}^*$ as an assumption in the cases where the plane reflection symmetry breaks down \cite{PhysRevD.89.061502,pan2014inspiral,PhysRevX.6.041014}.

In our SEOBNRE model, we follow the steps of SEOBNRv1 model to construct the radiation reaction force. The only difference is replacing the waveform with our SEOBNRE waveforms (\ref{Ewave}).

Besides the above method to calculate the energy flux, one may also calculate $\frac{dE}{dt}$ based on post-Newtonian approximation together with results from conservative part of EOBNR model. This is the method taken by ax-model \cite{PhysRevD.95.024038}. Similar to the idea we taken to treat the waveform in the above sub-section A, we divide the energy flux into two parts which correspond to the circular part and the non-circular correction part. Then the over all energy flux can be written as
\begin{align}
\frac{dE}{dt}=\frac{dE}{dt}|_{(C)}+\frac{dE}{dt}|_{Elip}-\frac{dE}{dt}|_{Elip,\dot{r}=0}.\label{energyflux}
\end{align}
We give the detail calculations for the post-Newtonian energy fluxes including $\frac{dE}{dt}|_{(C)}$ and $\frac{dE}{dt}|_{Elip}$ for eccentric binary in the Appendix~\ref{App:PNenergy}.
\begin{figure*}
\begin{tabular}{c}
\includegraphics[width=\textwidth]{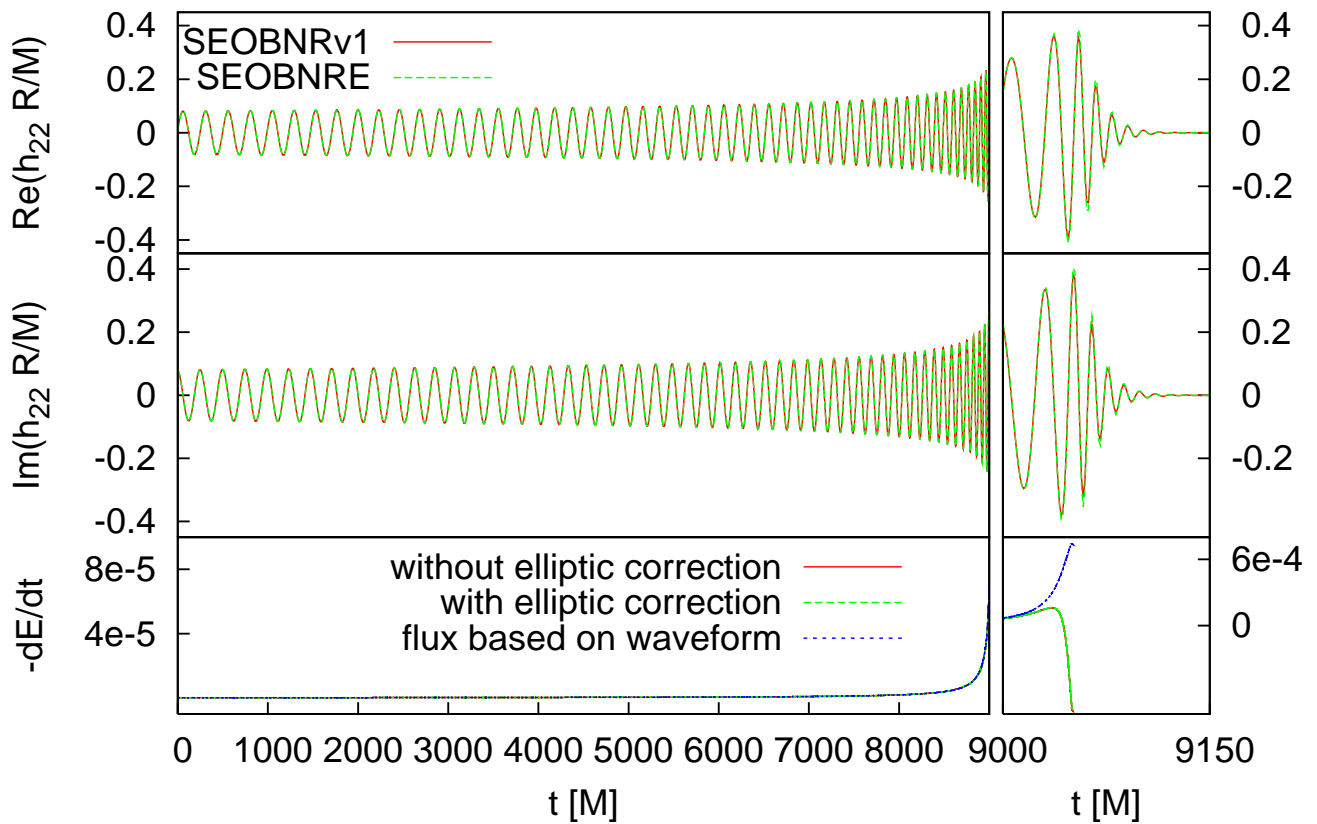}
\end{tabular}
\caption{Test for two identical spinless black holes with eccentricity $e_0=0$. The four top panels are comparison of ($\ell=2,m=2$) mode of the gravitational wave form for SEOBNRE and SEOBNRv1. The bottom two panels are comparison of energy flux for the one generated with waveform (\ref{waveformflux}), $-\frac{dE}{dt}$ given in (\ref{energyflux}) and $-\frac{dE}{dt}|_{(C)}$ in (\ref{energyfluxwo}). In the plot the energy flux generated with waveform (\ref{waveformflux}) is marked with `flux based on waveform'. The $-\frac{dE}{dt}$ given in (\ref{energyflux}) is marked with `with elliptic correction'. The $-\frac{dE}{dt}|_{(C)}$ in (\ref{energyfluxwo}) is marked with `without elliptic correction'. SEOBNRE model is used in the energy flux comparison.}\label{fig1}
\end{figure*}

\begin{figure*}
\begin{tabular}{c}
\includegraphics[width=\textwidth]{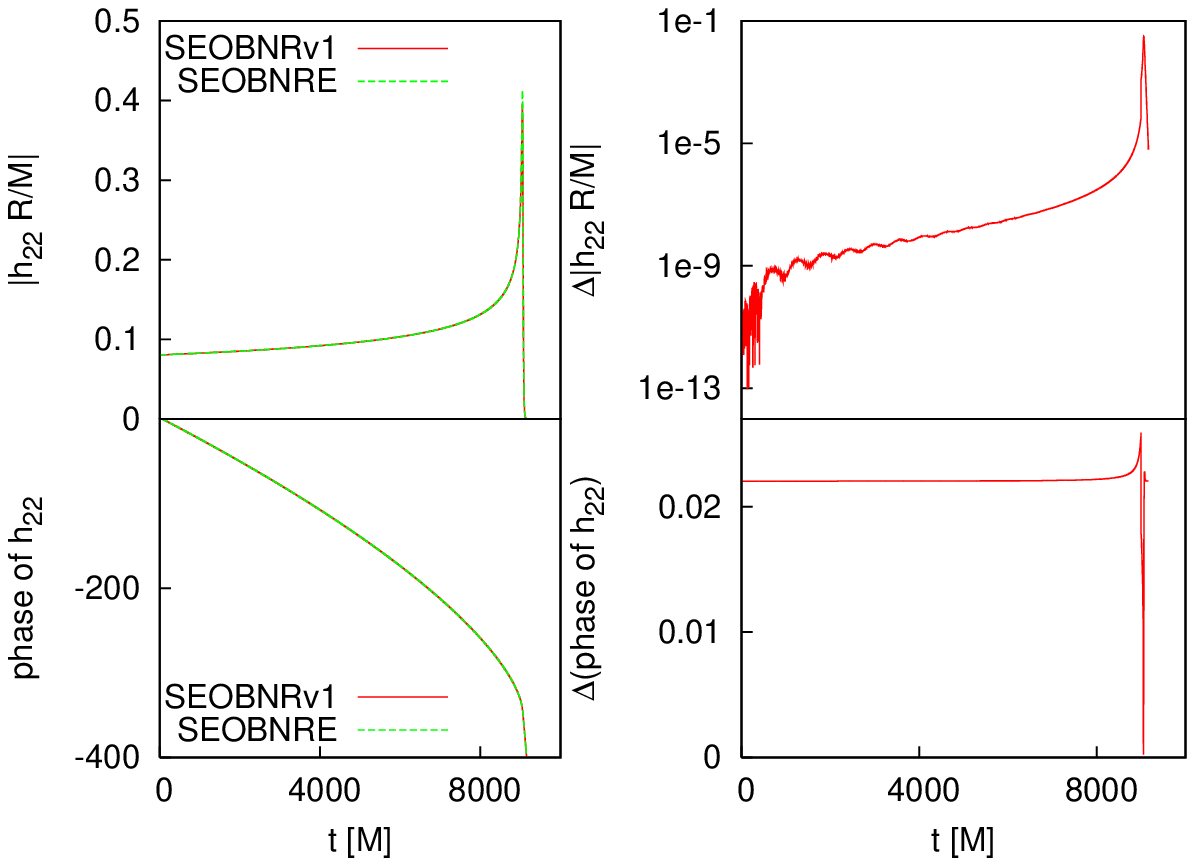}
\end{tabular}
\caption{Quantitative comparison of waveform $h_{22}$ corresponding to Fig.~\ref{fig1}. In the plot $\Delta |h_{22}R/M|\equiv \text{abs}(|h_{22,\text{SEOBNRE}}R/M| - |h_{22,\text{SEOBNRv1}}R/M|)$, $\Delta (\text{phase of }h_{22})\equiv\text{abs}(\text{phase of }h_{22,\text{SEOBNRE}}-\text{phase of }h_{22,\text{SEOBNRv1}})$.}\label{fig2}
\end{figure*}

\subsection{Initial data setting for the SEOBNRE dynamics}
Within EOB framework, we solve the dynamical equations (\ref{fdynr}) and (\ref{fdynp}), then plug the evolved dynamical variables into the waveform expression (\ref{Ewave}). But firstly we need to setup the initial value for the dynamical variables $(\vec{r},\vec{\tilde{p}})$. We take two steps to set the initial values. First, we look for the dynamical variable values for circular orbit (the authors in \cite{PhysRevD.74.104005} call it spherical orbit). Secondly, we adjust the momentum to achieve wanted eccentricity. In this work we consider binary black holes with spin perpendicular to the orbital plane. So the dynamics can be described with the test particle moves on the ecliptic plane of the central deformed Kerr geometry \cite{PhysRevD.84.124052}. Within the Boyer-Lindquist coordinate, we have $\phi=0,\theta=\frac{\pi}{2},\tilde{p}_\theta=0$. In order to get $r$, $\tilde{p}_r$ and $\tilde{p}_\phi$, we follow the Eqs.~(4.8) and (4.9) of \cite{PhysRevD.74.104005} to solve
\begin{align}
\frac{dH}{dr}&=0,\\
\frac{dH}{d\tilde{p}_\theta}&=0,\\
\frac{dH}{d\tilde{p}_\phi}&=\omega_0,
\end{align}
where $\omega_0=\frac{\pi}{f_0}$ is the speculated orbital frequency at initial time and the $f_0$ is the given frequency for gravitational wave at initial time. Assume the resulted solution for $r$ is $\hat{r}$, we adjust $r$ through
\begin{align}
r_0=\hat{r}/(1+e_0),
\end{align}
which means we put the test particle on the periastron of an elliptic orbit with eccentricity $e_0$ based on Newtonian picture \cite{han2014gravitational}. If the approximation of test particle and Newtonian picture is not good enough, our initial data setting method can not work well. More sophisticated initial conditions for eccentric binary black hole system are possible \cite{PhysRevD.89.084006}. We leave such investigations to future study.

\subsection{Match the inspiral waveform to the merger-ringdown waveform}
Like other EOBNR models, we assume the ringdown waveform can be described by the combination of quasi normal modes as
\begin{align}
h^{\text{merger-RD}}_{\ell m}=\sum_{n=0}^{N-1}A_{\ell mn}e^{-i\sigma_{\ell m n}(t-t^{\ell m}_{\text{match}})},\label{RDwaveform}
\end{align}
where $\sigma_{\ell m n}$ are the complex eigen values of the corresponding quasi normal modes for a Kerr black hole, $t^{\ell m}_{\text{match}}$ is the matching time point and $A_{\ell mn}$ are the combination coefficients for each mode. The same to SEOBNRv1 \cite{SEOBNRv1code} we take $N=8$.

In order to determine $\sigma_{\ell m n}$, we need to know the mass and spin of the final Kerr black hole. In principle the mass and the spin may be affected by the eccentricity. But here we neglect such dependence on the eccentricity as \cite{PhysRevD.95.024038} based on the assumption that the eccentricity is small for the cases considered in current work. We specify the mass and the spin of the final black hole following \cite{barausse2009predicting,PhysRevD.78.081501}
\begin{align}
&M_{\text{final}}=M[1+4(m^0-1)\eta+16m^1\eta^2(\chi_1+\chi_2)],\label{finalmass}\\
&\chi_{\text{final}}\equiv\frac{a_{\text{final}}}{M_{\text{final}}}=\chi^0+\eta\chi^0(t_4\chi^0+t_5\eta+t_0)\nonumber\\
&+\eta(2\sqrt{3}+t_2\eta+t_3\eta^2),\label{finalspin}\\
&m^0=0.9515,m^1=-0.013,\nonumber\\
&q=\frac{m_1}{m_2},\chi^0=\frac{\chi_1+\chi_2q^2}{1+q^2},\chi_{1,2}\equiv\frac{a_{1,2}}{m_{1,2}},\nonumber\\
&t_0=-2.8904,t_2=-3.5171,t_3=2.5763,\nonumber\\
&t_4=-0.1229,t_5=0.4537.\nonumber
\end{align}
Note that the above relations are valid only for the spins of the two black holes perpendicular to the orbital plane which are the cases considered in current work.

Regarding to the matching time point $t^{\ell m}_{\text{match}}$, we determine it based on inspiral dynamics as following. For the inspiral part, we solve the dynamics (\ref{fdynr}) and (\ref{fdynp}) till a time point which is called `merger time point' for convenience. The criteria of the `merger time point' is $r<6M$ and the orbital frequency begins to decrease. Then we chose the matching time point $t^{\ell m}_{\text{match}}$ corresponding to the time of the peak amplitude of the waveform $h^{insp-plun}_{\ell m}$.

At last we determine the coefficients $A_{\ell mn}$ base on the rule that the matching is smooth at $t^{\ell m}_{\text{match}}$ through first order derivative of the waveform.
\section{Test results for SEOBNRE model}\label{sec::tests}
In this section we will compare our SEOBNRE model to several existing waveform results including SEOBNR model, ax model, numerical relativity simulation and Teukolsky equation results. In addition to the comparison between the waveforms directly, we use overlap factor to quantify the difference between our SEOBNRE model and these existing waveform results \cite{PhysRevD.90.062003}. For two waveforms $h(t)$ and $s(t)$, the overlap factor is defined as
\begin{align}
\mathcal{O}(h,s)\equiv\frac{\langle h|s\rangle}{\sqrt{\langle h|h\rangle\langle s|s\rangle}},\label{overlapfactordef}
\end{align}
with the inner product defined as
\begin{align}
\langle h|s\rangle\equiv4Re\int_{f_{min}}^{f_{max}}\frac{\tilde{h}(f)\tilde{s}^*(f)}{S_n(f)}df
\end{align}
where $\tilde{}$ means the Fourier transformation, $f$ represents frequency, $S_n(f)$ is the one sided power spectral density of the detector noise, and ($f_{min}$, $f_{max}$) is the frequency range of the detector. As a typical example, we consider advanced LIGO detector in the following investigations. More specifically, we take the sensitivity of LIGO-Hanford during O1 run as our $S_n(f)$ which is got from the LSC webpage \cite{LSC150914}. Correspondingly we take $f_{min}=20$Hz and $f_{max}=2000$Hz. Like other existing waveform models, the total mass $M$ of the binary black hole, the source location ($\theta$, $\phi$), the angles between the eccentric orbit and the line direction ($\iota$, $\beta$) and the polarization angle $\psi$ are free parameters \cite{PhysRevD.92.044034}. But in order to let the waveform falls in LIGO's frequency band, we choose $M=20$M$_\odot$ as an example to calculate the overlap factor $\mathcal{O}$. Regarding to the five angles, values $\theta=\phi=\iota=\beta=\psi=0$ are taken.
\begin{figure}
\begin{tabular}{c}
\includegraphics[width=0.45\textwidth]{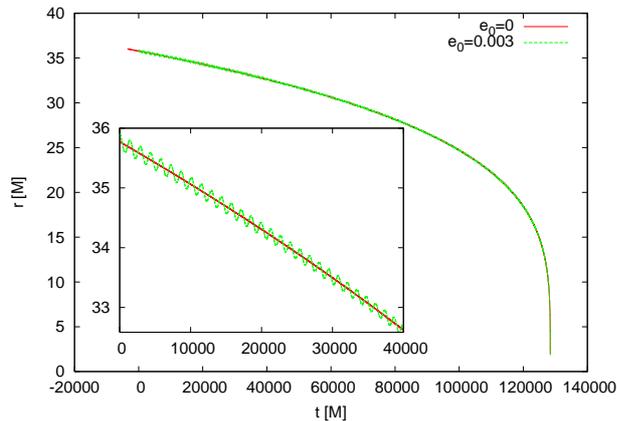}
\end{tabular}
\caption{Test for two identical spinless black holes with eccentricity $e_0=0.003$ at $Mf_0\approx0.0015$. The comparison of orbital evolution for $e_0=0.003$ and $e_0=0$. The result of $e_0=0$ is got through SEOBNRv1 code, and the $e_0=0.003$ result is got by SEOBNRE model. The inset of the plot is the blowup of the corresponding range.}\label{fig3}
\end{figure}

\subsection{Comparison to SEOBNRv1}
Firstly we compare the result of SEOBNRE model with $e_0=0$ against to SEOBNRv1. We consider two spinless black holes with equal mass. In Fig.~\ref{fig1} $t^{22}_{\text{match}}\approx9048.1579M$. The overlap factor between the two waveforms shown in Fig.~\ref{fig1} is $\mathcal{O}=0.99998$. At the same time we have also checked the energy flux $-\frac{dE}{dt}|_{(C)}$ introduced in \cite{PhysRevD.95.024038} which is shown in (\ref{energyfluxwo}) and the elliptical orbit correction we calculated in (\ref{energyflux}). We find that the energy flux (\ref{energyfluxwo}) can describe the real flux used by SEOBNRE dynamics (\ref{waveformflux}) quite well in the early inspiral stage but fails at late inspiral and plunge stage. At later times, the energy flux (\ref{energyfluxwo}) even becomes positive which is unphysical. This implies that the PN energy flux expression breaks down at late inspiral and plunge stage.

In Fig.~\ref{fig2} we compare the waveform $h_{22}$ more quantitatively, where the amplitude and phase are considered. Regarding to the elliptical orbit correction terms, as ones expect, they are ignorable before plunge in this quasi-circular case as shown in the bottom panel of Fig.~\ref{fig1}. Near merger, our elliptic correction terms fail to distinguish real eccentric orbit with the plunge behavior in the quasi-circular orbit, so the difference for both amplitude and phase increase. At merger, the differences for amplitude and phase get maximal values, about 0.03 and 0.026rad respectively. As shown in Fig.~\ref{fig1} and Fig.~\ref{fig2}, our SEOBNRE model can recover SEOBNRv1 result some well. In Fig.~\ref{fig1} and Fig.~\ref{fig2} we align the time at simulation start time ($t=0$) which corresponds to gravitational wave frequency $Mf_0\approx0.004$. For two 10M${}_\odot$ black holes $f_0=40$Hz. At this alignment time, we also set the phase of the gravitational wave 0 which makes the comparison easier.

\begin{figure*}
\begin{tabular}{c}
\includegraphics[width=\textwidth]{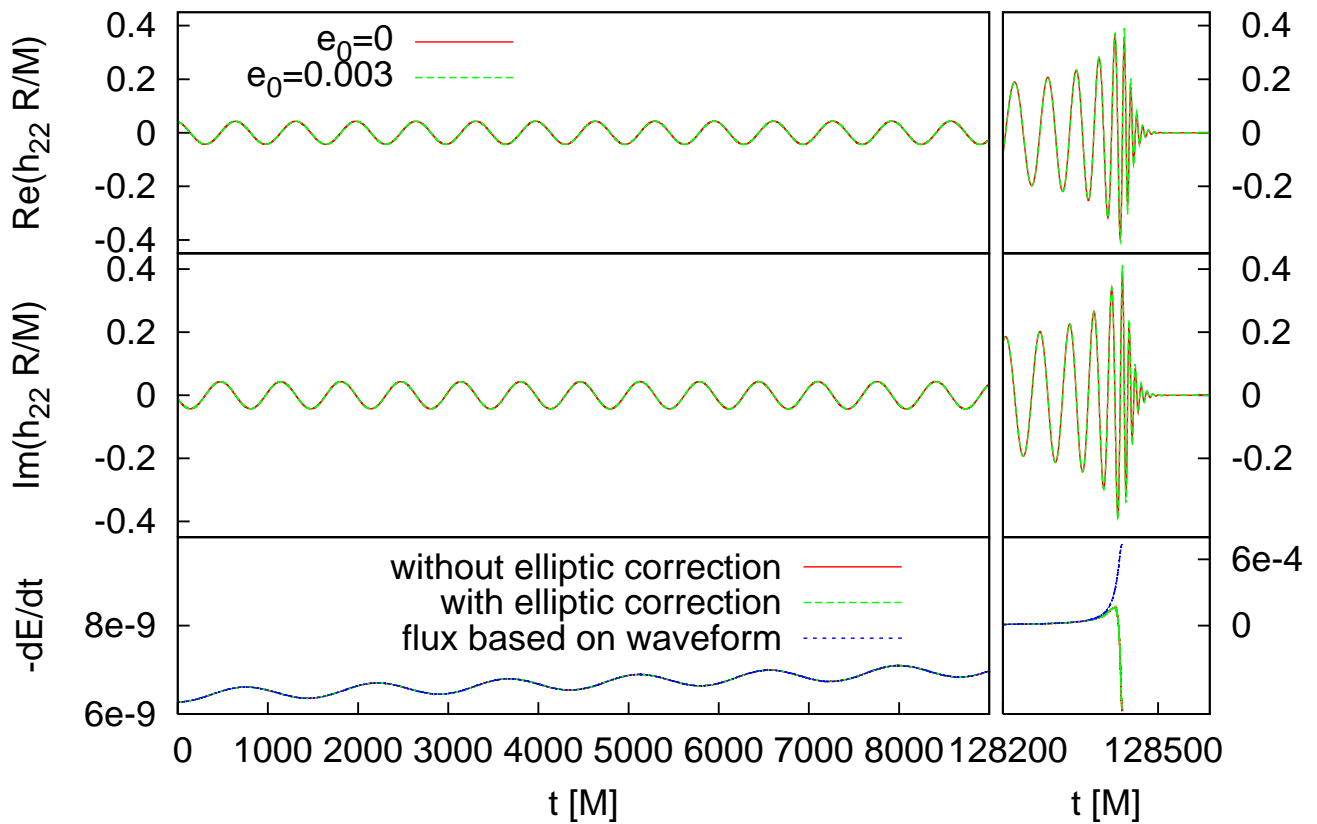}
\end{tabular}
\caption{Similar to Fig.~\ref{fig1} but for two identical spinless black holes with eccentricity  $e_0=0.003$ at $Mf_0\approx0.0015$. The result of $e_0=0$ is got through SEOBNRv1 code, and the $e_0=0.003$ result is got by SEOBNRE model.}\label{fig4}
\end{figure*}

Our second testing case is two identical spinless black holes with eccentricity $e_0=0.003$ at $Mf_0\approx0.001477647$ which corresponds to two 10M${}_\odot$ black holes with $f_0=15$Hz. As shown in Fig.~\ref{fig3} we can see clearly the oscillation of radial coordinate $r$ respect to time which corresponds to the eccentric orbital motion. But this level of eccentricity is ignorable for waveform as shown in Fig.~\ref{fig4}. Although some small, we can see the oscillation behavior of the energy flux with the same frequency to the $r$ motion in the bottom panel of Fig.~\ref{fig4}. And again we find that the energy flux (\ref{energyfluxwo}) can describe the real flux quite well in the early inspiral stage but fails at late inspiral and plunge stage. In Fig.~\ref{fig3} and Fig.~\ref{fig4}, we have adjusted the time coordinate of $e_0=0$ result to align with the merger time of $e_0=0.003$ which is $t^{22}_{\text{match}}\approx128427.86M$.
The overlap factor between the two waveforms shown in Fig.~\ref{fig4} is $\mathcal{O}=0.99995$.

\begin{figure*}
\begin{tabular}{c}
\includegraphics[width=\textwidth]{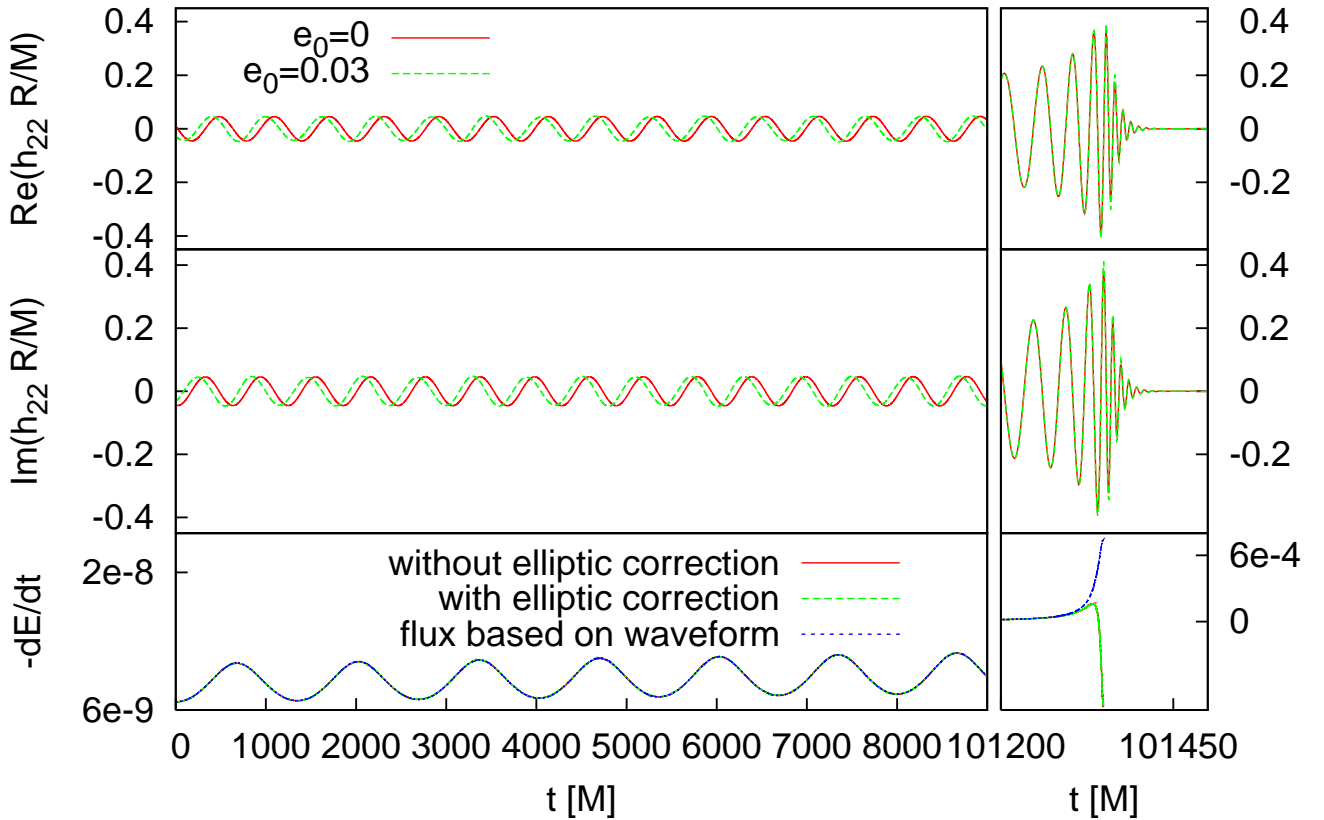}
\end{tabular}
\caption{Similar to Fig.~\ref{fig1} but for two identical spinless black holes with eccentricity  $e_0=0.03$ at $Mf_0\approx0.0015$. The result of $e_0=0$ is got through SEOBNRv1 code, and the $e_0=0.03$ result is got by SEOBNRE model.}\label{fig5}
\end{figure*}

\begin{figure*}
\begin{tabular}{c}
\includegraphics[width=\textwidth]{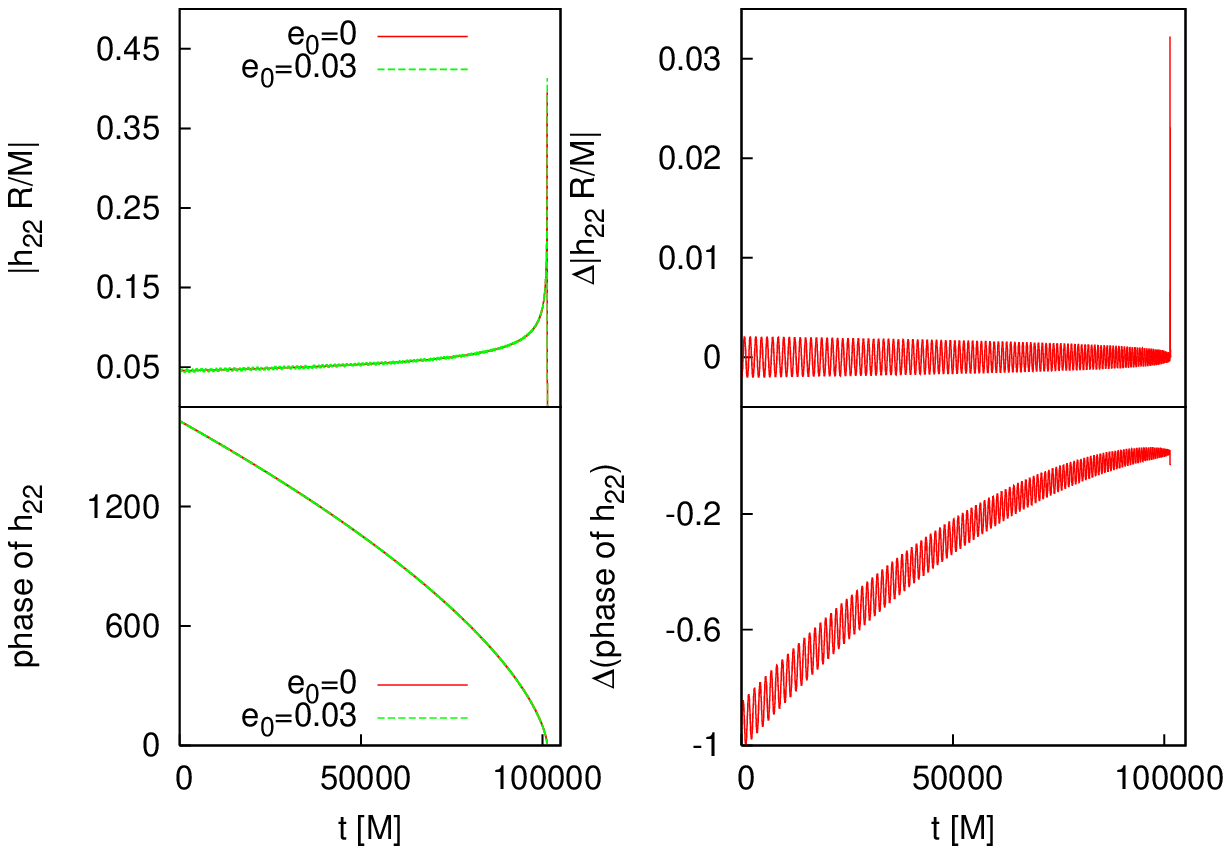}
\end{tabular}
\caption{Amplitude and phase comparison for $h_{22}$ corresponding to the case $e_0=0.03$ shown in Fig.~\ref{fig5}. The result of $e_0=0$ is got through SEOBNRv1 code, and the $e_0=0.03$ result is got by SEOBNRE model.}\label{fig6}
\end{figure*}

When we increase the eccentricity to $e_0=0.03$ for $Mf_0\approx0.001477647$, the radial oscillation becomes stronger than that of Fig.~\ref{fig3}. But the overall behavior is similar. Regarding to the waveform, phase difference to quasi-circular case appears as shown in Fig.~\ref{fig5}. More quantitatively the amplitude difference and the phase difference are shown in Fig.~\ref{fig6}. Along with the time, the two differences decrease. This is because the eccentricity is decreasing due to the circularizing effect of gravitational radiation. Near merger, such difference almost disappears. This result supports our assumption that we ignore the effect of eccentricity on the mass and spin of the final Kerr black hole in (\ref{finalmass}) and (\ref{finalspin}). Near merger the difference becomes larger again. But we note that the maximal difference is 0.03 and 0.02rad for amplitude and phase respectively. This level of difference is the same to the one we got for quasi-circular case shown in Fig.~\ref{fig2}. So we believe this is resulted from the same reason as in quasi-circular case. Similar to Fig.~\ref{fig4}, we here have adjusted the time coordinate of $e_0=0$ result to align with the merger time of $e_0=0.03$ which is $t^{22}_{\text{match}}\approx101345.99M$. The overlap factor between the two waveforms shown in Fig.~\ref{fig5} is $\mathcal{O}=0.99300$.

\begin{figure*}
\begin{tabular}{c}
\includegraphics[width=\textwidth]{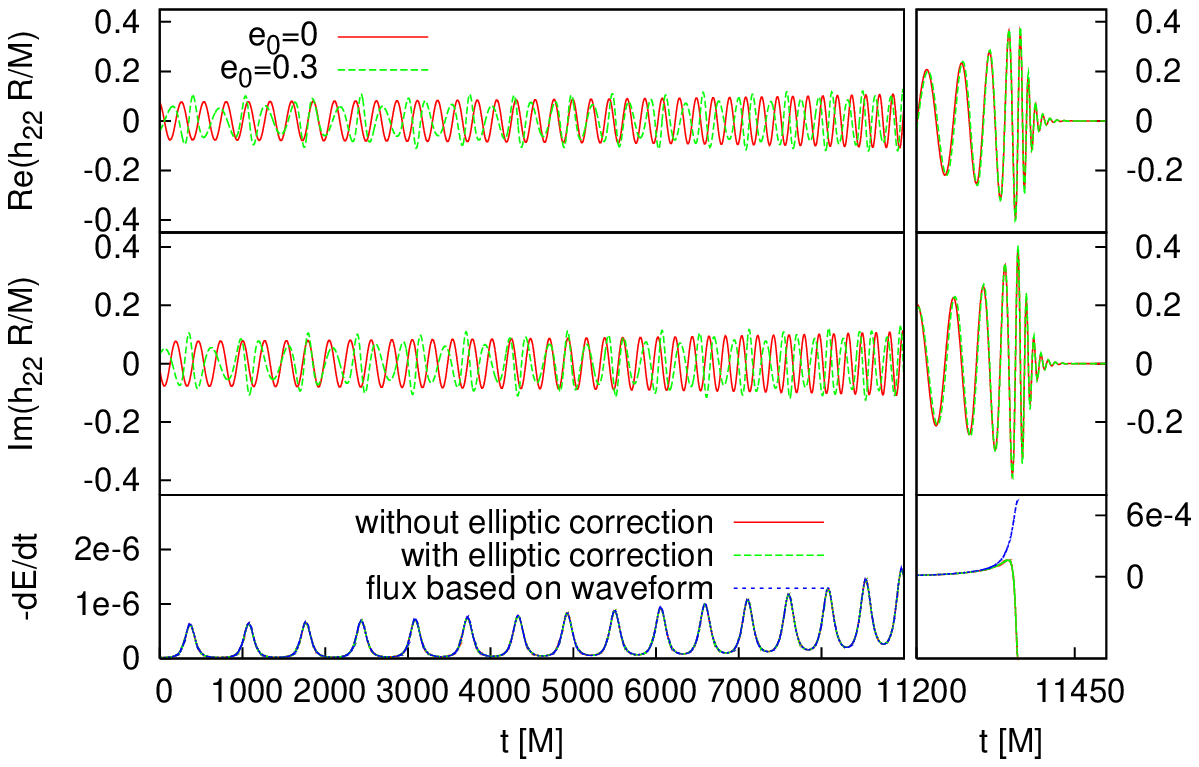}
\end{tabular}
\caption{Similar to Fig.~\ref{fig1} but for two identical spinless black holes with eccentricity $e_0=0.3$ at $Mf_0\approx0.0015$. The result of $e_0=0$ is got through SEOBNRv1 code, and the $e_0=0.3$ result is got by SEOBNRE model.}\label{fig7}
\end{figure*}

\begin{figure*}
\begin{tabular}{c}
\includegraphics[width=\textwidth]{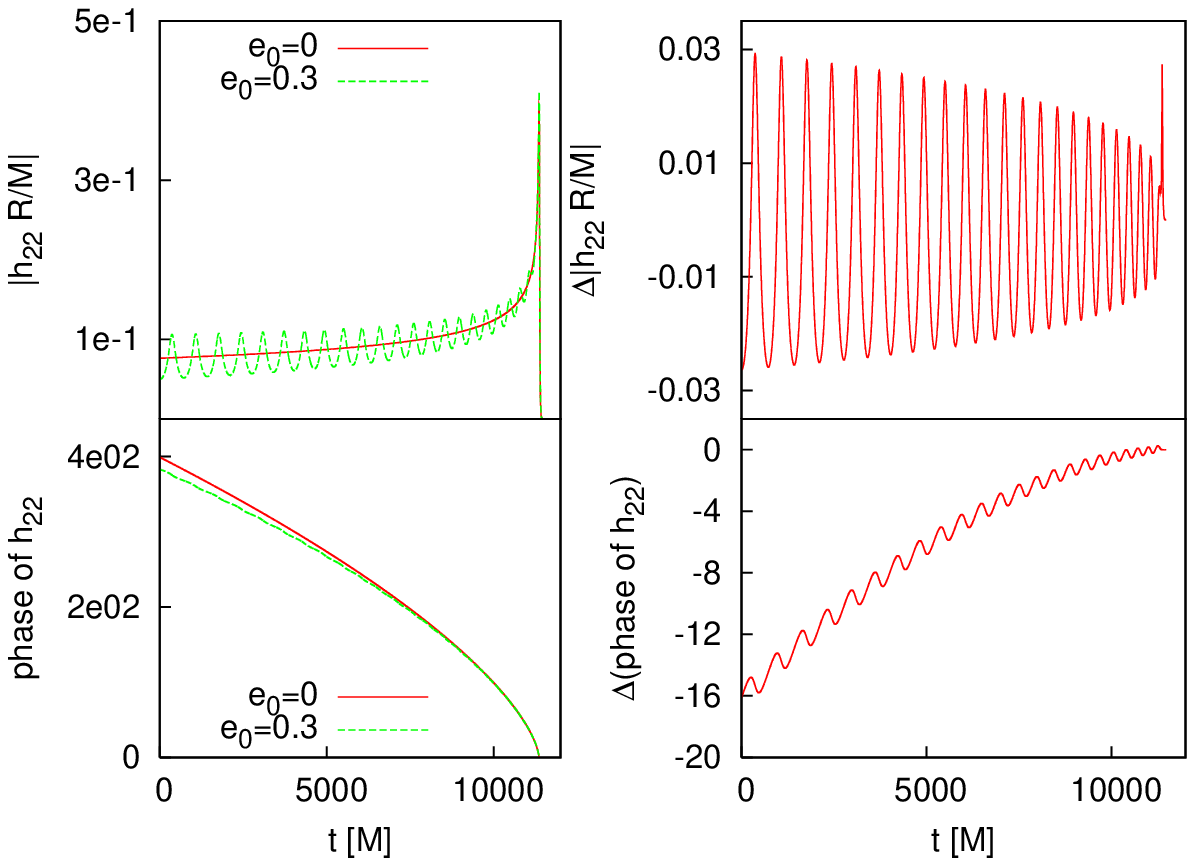}
\end{tabular}
\caption{Amplitude and phase comparison for $h_{22}$ corresponding to the case shown in Fig.~\ref{fig7}. The result of $e_0=0$ is got through SEOBNRv1 code, and the $e_0=0.3$ result is got by SEOBNRE model.}\label{fig8}
\end{figure*}

When we increase more the eccentricity to $e_0=0.3$ for $Mf_0\approx0.001477647$, the radial oscillation becomes much stronger as expected while the overall behavior is similar to that of Fig.~\ref{fig3}. Regarding to the waveform, the oscillation behavior of the amplitude can be clearly seen as shown in Fig.~\ref{fig7}. The amplitude difference and the phase difference are quantitatively shown in Fig.~\ref{fig8}. The overall behavior is similar to that of Fig.~\ref{fig6}. In this case, the maximal difference in the early inspiral stage is 0.03 and 16rad for amplitude and phase respectively. Similar to Fig.~\ref{fig4}, we here have adjusted the time coordinate of $e_0=0$ result to align with the merger time of $e_0=0.3$ which is $t^{22}_{\text{match}}\approx11357.098M$. The overlap factor between the two waveforms shown in Fig.~\ref{fig7} is $\mathcal{O}=0.46942$.

When the initial eccentricity $e_0$ becomes bigger than $0.6$ for $Mf_0\approx0.001477647$, we find that the correction term of the waveform (\ref{h22willelip}) becomes significant which means the perturbation assumption of small eccentricity breaks down. This situation may be improved by higher order post-Newtonian result for eccentric orbit binary. But unfortunately the higher order PN results for eccentric orbit binary is not available yet. Of course, this does not mean our model can be applied to $e_0<0.6$ cases. Cases with such large eccentricity need more tests against to, for example, numerical relativity simulations. The tests done in this subsection indicate that our SEOBNRE model can give consistent results compared to quasi-circular case.

\subsection{Comparison to advanced x-model (ax model)}
In \cite{PhysRevD.82.024033} the authors proposed x-model to describe the gravitational waveform for eccentric orbital binary black hole. For the early inspiral stage, the authors of \cite{PhysRevD.82.024033} showed consistence between the x-model and the numerical relativity simulation result. Noting that the x-model is based on low order post-Newtonian approximation, Huerta and his coworkers improved x-model with high order PN results to get advanced x-model (ax model) in \cite{PhysRevD.95.024038}. The full ax model includes inspiral, merger and ringdown stages. Similar to our treatment about merger and ringdown, ax model also takes the assumption that the eccentricity is small and has been dissipated away before merger. So here we only compare our SEOBNRE model to ax model for inspiral part.

In ax model, the adiabatic picture is taken. So the eccentricity itself is treated as a dynamical variable. In all, ax model includes dynamical variables eccentricity $e$, reduced orbital frequency $x\approx(\omega M)^2/3$ with $\omega$ the orbital frequency, the mean anomaly $l$ and the relative angular coordinate $\phi$. In the comparison here, we take initial data $x=0.05$, $l=\phi=0$ and vary $e$.

\begin{figure*}
\begin{tabular}{c}
\includegraphics[width=\textwidth]{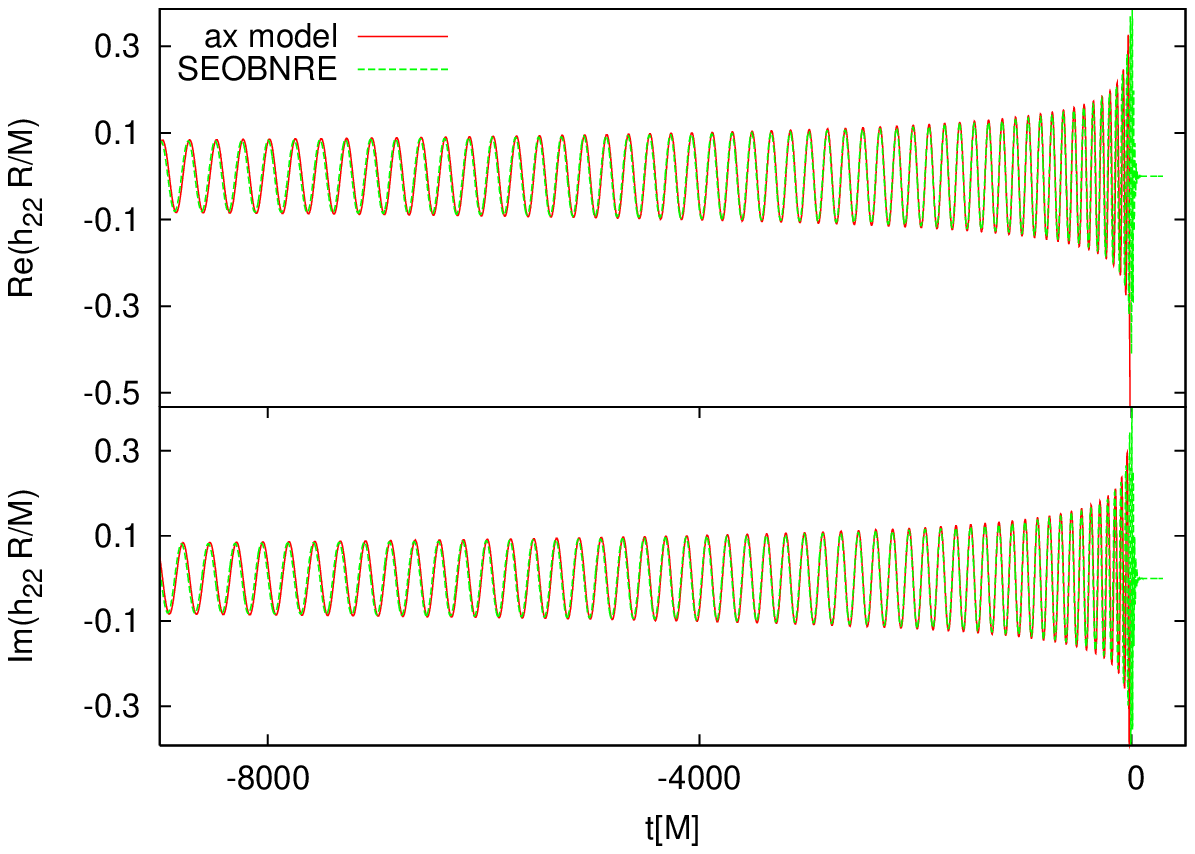}
\end{tabular}
\caption{Comparison between ax model and SEOBNRE model for $e_0=0$ case.}\label{fig9}
\end{figure*}

Firstly we compare the $e_0=0$ case in Fig.~\ref{fig9}. Overall the two waveforms show very good consistence which corresponds to the result of fitting factor $0.95$ respect to advanced LIGO got in \cite{PhysRevD.95.024038}. But we can still see some amplitude difference near merger, and some phase difference when the evolution time becomes longer. Since we have confirmed our SEOBNRE model in Fig.~\ref{fig1} and Fig.~\ref{fig2} for $e_0=0$ case, we attribute this difference to the relatively low order PN approximation of ax model. Effectively the EOBNR model admits more than 3.5 PN order \cite{PhysRevD.76.104049,messina2017back}, this is why we call ax model relatively low PN order. The overlap factor between the two waveforms (inspiral part, $t<0$) shown in Fig.~\ref{fig9} is $\mathcal{O}=0.99802$.

\begin{figure*}
\begin{tabular}{c}
\includegraphics[width=\textwidth]{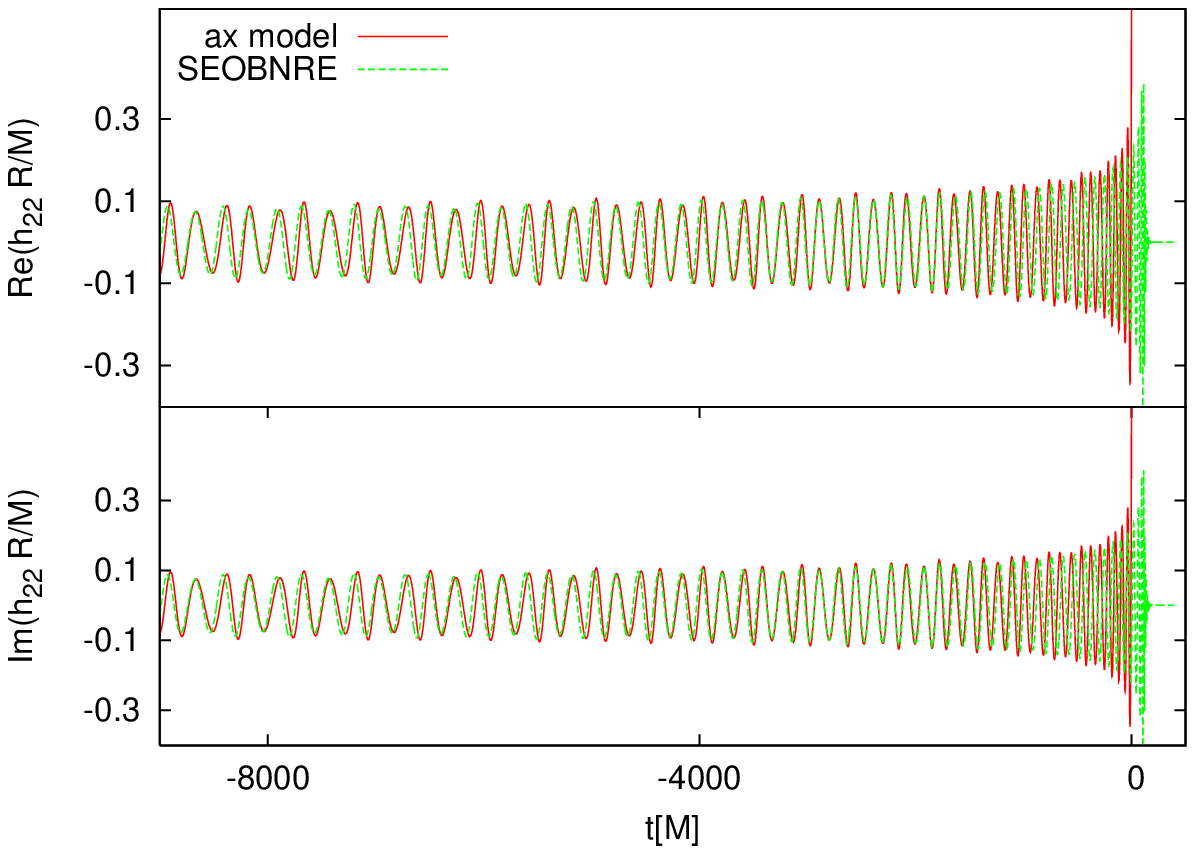}
\end{tabular}
\caption{Comparison between ax model ($e_0=0.1$, $x_0=0.05$) and SEOBNRE model ($e_0=0.15$, $Mf_0\approx0.001477647$).}\label{fig10}
\end{figure*}

\begin{figure*}
\begin{tabular}{c}
\includegraphics[width=\textwidth]{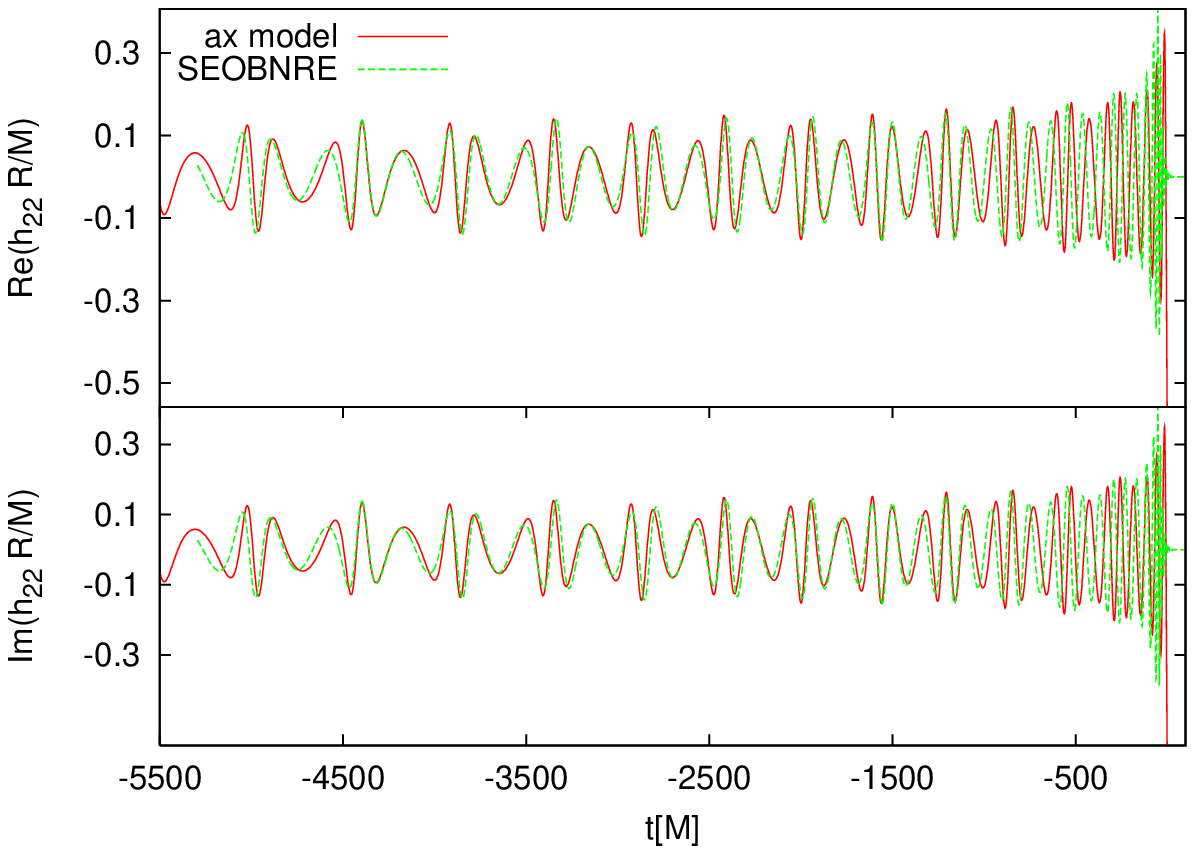}
\end{tabular}
\caption{Comparison between ax model ($e_0=0.3$, $x_0=0.05$) and SEOBNRE model ($e_0=0.4$, $Mf_0\approx0.001477647$).}\label{fig11}
\end{figure*}

In contrast to ax model, the eccentricity is not an explicit dynamical variable in SEOBNRE model. And limited by the simplified method we taken to treat the initial data described in the above section, we start our SEOBNRE simulation from somewhat low frequency $Mf_0\approx0.001477647$ and vary different $e_0$ to fit ax model result. The fitting process has not been optimized, so the real consistent result may be better than the ones presented here. We have considered two concrete examples. The first example is $e_0=0.1$ at $x_0=0.05$ for ax model and $e_0=0.15$ at $Mf_0\approx0.001477647$ for SEOBNRE model. The initial eccentricity for SEOBNRE model is larger. We suspect this is because the effective initial separation implemented in SEOBNRE model is larger than the one in ax model. After some evolution time, the circularization effect drives the eccentricity within SEOBNRE model to the one presented in ax model. The comparison result is shown in Fig.~\ref{fig10}. We can see the consistence of amplitude and phase between ax model and SEOBNRE model lasts more than 8000M. The overlap factor between the two waveforms shown in Fig.~\ref{fig10} is $\mathcal{O}=0.93031$. Our second example is the comparison between $e_0=0.3$ at $x_0=0.05$ of ax model and $e_0=0.4$ at $Mf_0\approx0.001477647$ of SEOBNRE model. The result is shown in Fig.~\ref{fig11}. The consistence of amplitude and phase between ax model and SEOBNRE model lasts about 5000M. The overlap factor between the two waveforms shown in Fig.~\ref{fig11} is $\mathcal{O}=0.66617$.

\subsection{Comparison to numerical relativity results}
No matter how reasonable, we have taken several approximations when we construct our SEOBNRE model. In contrast to this situation, numerical relativity solves Einstein equation directly \cite{baumgarte2010numerical}. Up to the numerical error, the results given by numerical relativity is the exact solution to the Einstein equation. So we can use the simulation results by numerical relativity to check and calibrate the validity of SEOBNRE model. Since the EOBNR models have been well calibrated to numerical relativity results for circular cases, and our SEOBNRE model can recover the usual EOBNR model as shown in Fig.~\ref{fig1} and Fig.~\ref{fig2} for $e_0=0$ case, there is no surprising that our SEOBNRE model is consistent to quasi-circular simulation results of numerical relativity.

For eccentric cases, there are some subtleties in defining the eccentricity due to the `in'-spiral effect reduced by the gravitational radiation. In this work we are not intent to touch this subtle problem involved in numerical relativity \cite{pfeiffer2007reducing,PhysRevD.82.124016}. Instead, we take the similar recipe adopted in the above subsection to do the comparison. For a given numerical relativity simulation result, we vary the initial eccentricity $e_0$ corresponding to $Mf_0\approx0.001477647$ for SEOBNRE to fit the numerical relativity result. Again the fitting process is not optimized, so the consistence between our SEOBNRE model and the numerical relativity result may be better than the ones presented here.

\begin{figure*}
\begin{tabular}{c}
\includegraphics[width=\textwidth]{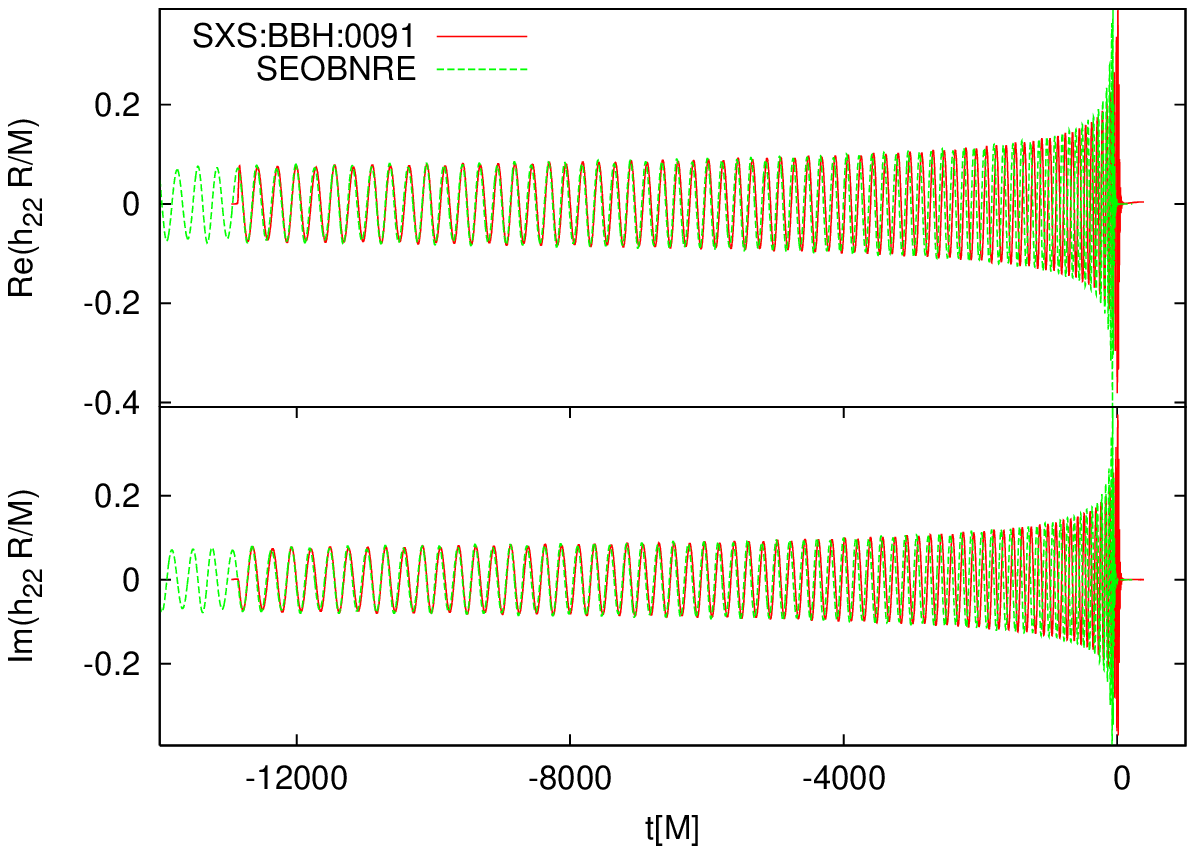}
\end{tabular}
\caption{SXS:BBH:0091 one is the SpEC simulation result for equal mass spinless binary black hole with $e_0=0.02181$ at orbital frequency $0.0105565727235$. SEOBNRE one corresponds to $e_0=0.1$ at $Mf_0\approx0.001477647$ for equal mass spinless binary black hole.}\label{fig12}
\end{figure*}

\begin{figure*}
\begin{tabular}{c}
\includegraphics[width=\textwidth]{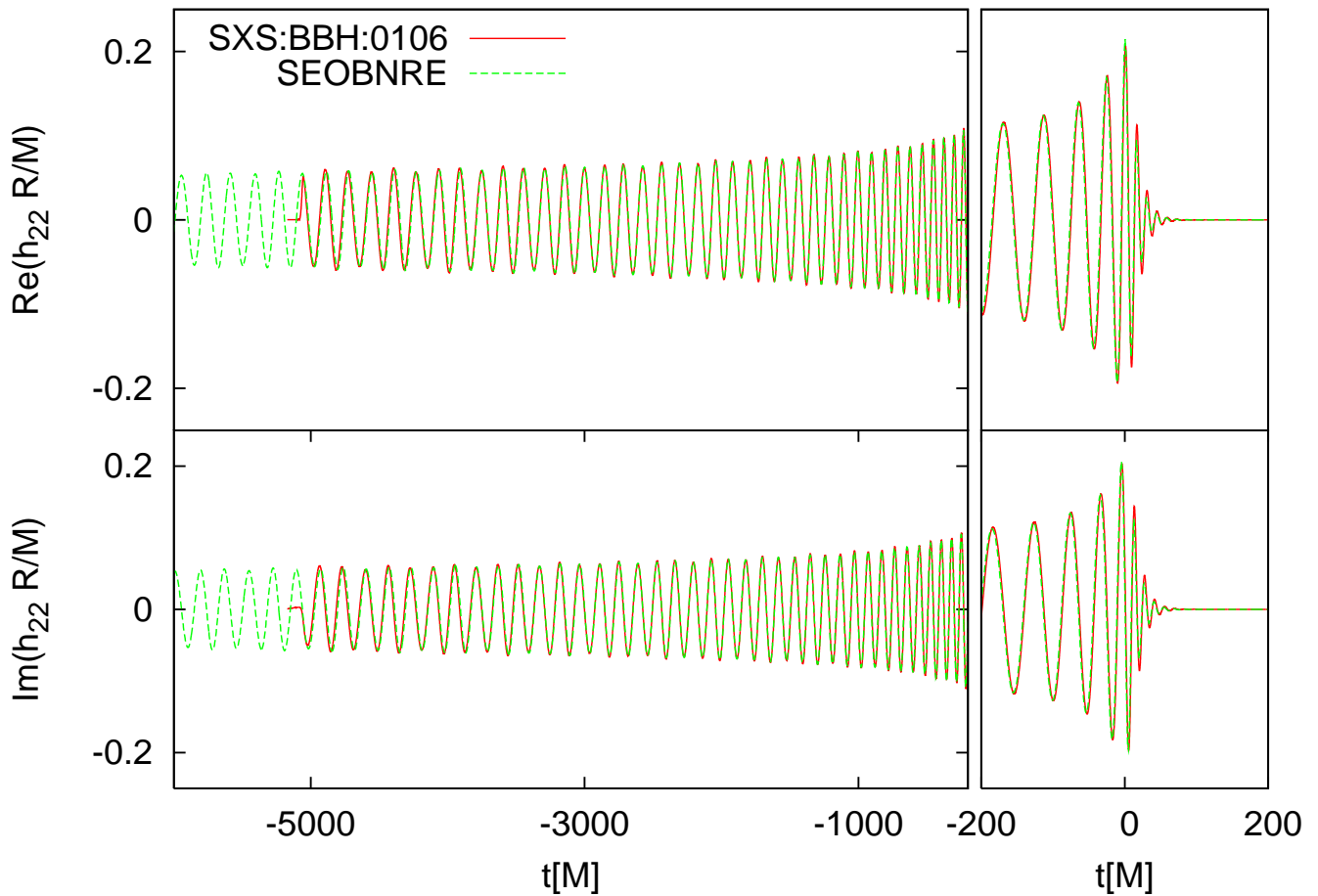}
\end{tabular}
\caption{SXS:BBH:0106 one is the SpEC simulation result for two spinless black holes with mass ratio $1:5$ start orbit with $e_0=0.02181$ at orbital frequency $0.0105565727235$. SEOBNRE one corresponds to $e_0=0.1$ at $Mf_0\approx0.001477647$ for two spinless black holes with mass ratio $1:5$.}\label{fig13}
\end{figure*}

We have done three comparisons in the current work. The numerical relativity simulation results come from the public data \cite{SXSBBH} which are calculated by SpEC code \cite{PhysRevD.86.084033}. The first one is the waveform SXS:BBH:0091 \cite{SXSBBH} which corresponds to an equal mass, spinless binary black hole with initial eccentricity $e_0=0.02181$ starting evolve at orbital frequency $Mf_0=0.0105565727235$. On the SEOBNRE side, the initial eccentricity is $e_0=0.1$ starting evolve at orbital frequency $Mf_0=0.001477647$ which is the same as the ones shown in previous figures. The comparison is presented in Fig.~\ref{fig12}. Although the eccentricity involved in this comparison is some small, the oscillation of the gravitational waveform amplitude is clear. The consistence for both the amplitude and the phase is quite good. The overlap factor between the two waveforms shown in Fig.~\ref{fig12} is $\mathcal{O}=0.99201$. At the same time we also note that the overlap factor between the numerical relativity waveform and the $e_0=0$ SEOBNRE waveform is 0.990. Our second comparison for numerical relativity simulation result is SXS:BBH:0106 \cite{SXSBBH}. Both the numerical relativity simulations result and the SEOBNRE simulation admit exactly the same eccentricity parameters as the first comparison. The only difference to the first comparison is the mass ratio for the binary black hole which is $5:1$ here. The comparison is shown in Fig.~\ref{fig13}. The overlap factor between the two waveforms shown in Fig.~\ref{fig13} is $\mathcal{O}=0.99739$. In contrast, the overlapping factor between the numerical relativity waveform and the $e_0=0$ SEOBNRE waveform is 0.989. Interestingly we find that the consistence between numerical relativity result and SEOBNRE result is even better than the first one. The SEOBNRE result can recover the numerical relativity result for the whole inspiral-merger-ringdown process. We can understand this result as following. As shown by Peters through post-Newtonian approximation in \cite{PhysRev.136.B1224}, the decay of eccentricity and the lifetime for the binary can be estimated
\begin{align}
\frac{de}{dt}&=-\frac{304}{15}\frac{M^3\eta}{a^4(1-e^2)^{5/2}}e(1+\frac{121}{304}e^2),\\
T(a_0,e_0)&=\frac{768}{425}\frac{5a_0^4}{256M^3\eta}(1-e_0^2)^{7/2},
\end{align}
where $a$ means the separation (semimajor axis) of the binary, and the subindex $0$ means the initial quantities. Definitely this estimation can not be correct for the late inspiral and merger stages considered in current paper. But this estimation can give us a qualitative picture. Compare the Fig.~\ref{fig12} and the Fig.~\ref{fig13}, we can see the lifetime for mass ratio $5:1$ binary is much shorter than the one for equal mass case. According to the above lifetime estimation, we can deduce that the initial separation for mass ratio $5:1$ binary is shorter than that of equal mass one. The initial separations used in the numerical relativity simulations are $19$ for equal mass binary and $14$ for mass ratio $5:1$ binary respectively. Although these separation values are gauge dependent, they show consistence to the PN prediction. Then note that the eccentricity decay is proportional to the fourth power of $a$, we can expect that the eccentricity decay involved in Fig.~\ref{fig13} case is much faster than that in Fig.~\ref{fig12}. Faster eccentricity decay results in smaller eccentricity during the later process. So our small eccentricity assumption works better.

\begin{figure*}
\begin{tabular}{c}
\includegraphics[width=\textwidth]{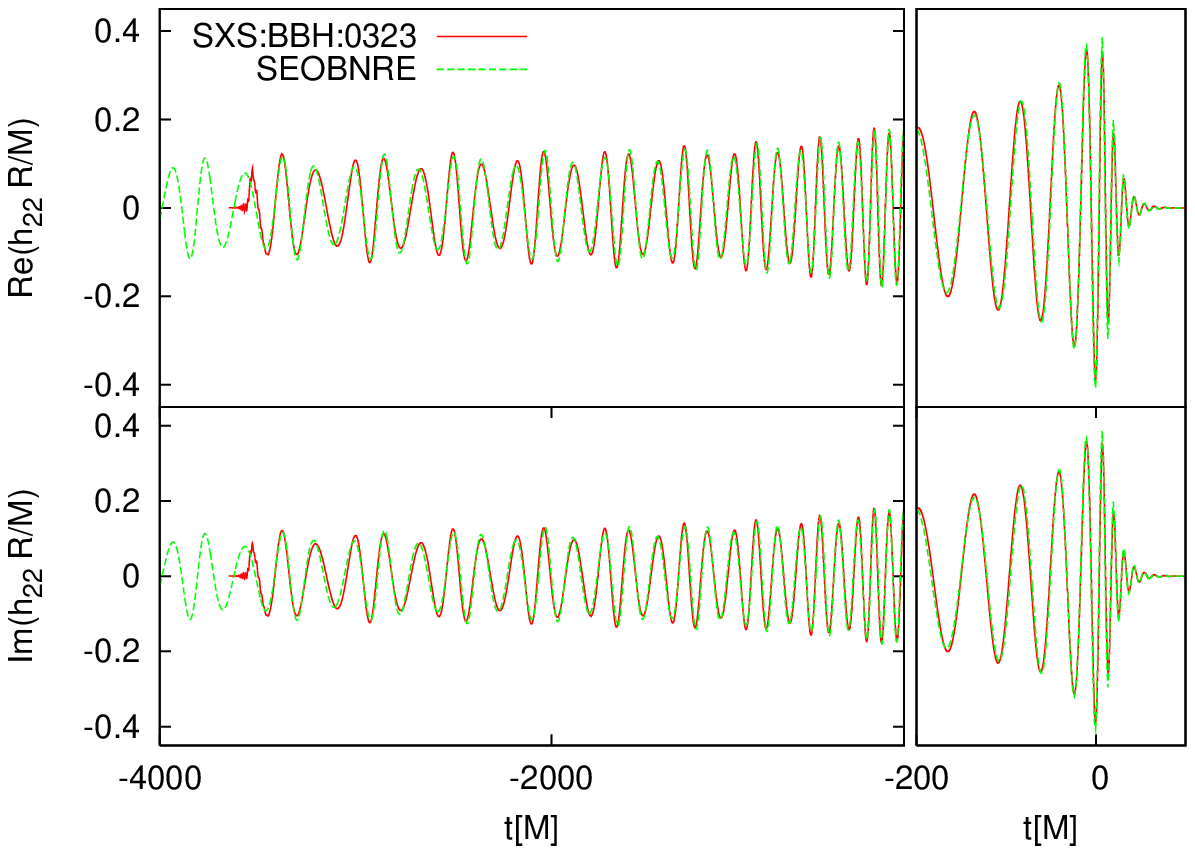}
\end{tabular}
\caption{SXS:BBH:0323 one is the SpEC simulation result for two spinless black holes with equal mass start orbit with $e_0=0.1935665$ at orbital frequency $0.0146842176288$. SEOBNRE one corresponds to $e_0=0.3$ at $Mf_0\approx0.001477647$ for two black holes with mass ratio 11:9 and spin $\chi_1=0.33$ and $\chi_2=-0.44$ respectively which corresponds to the setting of SXS:BBH:0323.}\label{fig14}
\end{figure*}

Our third comparison investigates a larger eccentricity case for equal mass binary. On the numerical relativity simulation side, the initial eccentricity is $e_0=0.1935665$ starting evolve at orbital frequency $Mf_0=0.0146842176288$. The eccentricity is about one order larger than the above two cases. The simulation data corresponds to SXS:BBH:0323 \cite{SXSBBH}. The mass ratio of the two black holes in this simulation is 11:9. And the dimensionless spins for the big and small black hole are $0.33$ and $-0.44$ respectively. On the SEOBNRE side, the initial eccentricity is $e_0=0.3$ starting evolve at orbital frequency $Mf_0=0.001477647$. The comparison is shown in Fig.~\ref{fig14}. During the inspiral stage, we can see the waveform amplitude and phase are roughly consistent between the numerical relativity result and the SEOBNRE result. The overlap factor between the two waveforms shown in Fig.~\ref{fig14} is $\mathcal{O}=0.98171$. In contrast, the overlap factor between the numerical relativity waveform and the $e_0=0$ SEOBNRE waveform is 0.849. Due to the larger eccentricity, the consistence is not as good as $e_0\approx0.02$ cases shown in Fig.~\ref{fig12} and Fig.~\ref{fig13}. But we can note that the consistence is quite good for merger and ringdown stage. We suspect this is because the eccentricity has decayed quite amount, so the consistence improves during these stages.

\subsection{Comparison to Teukolsky equation results for extreme mass ratio binary}
\begin{figure*}
\begin{tabular}{c}
\includegraphics[width=\textwidth]{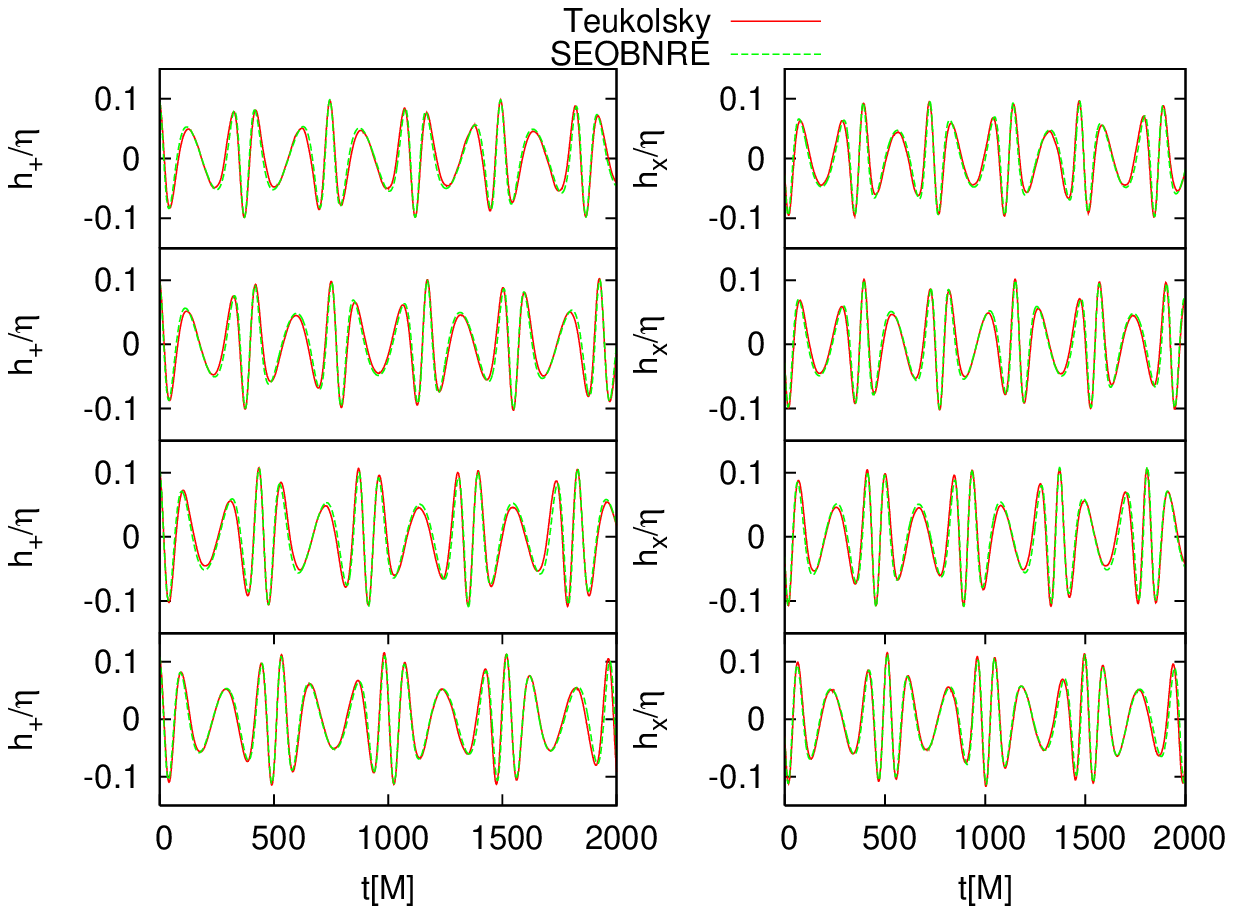}
\end{tabular}
\caption{Comparison between Teukolsky equation based model (marked with `Teukolsky') and SEOBNRE model. All cases here admit mass ratio about $1:1000$, initial eccentricity $e_0=0.3$, and the initial separation parameter $p_0=12$. From top row to bottom row, the corresponding initial orbital angular momentum are $\chi=0.9$, $p_{\phi_0}=3.75$; $\chi=0.5$, $p_{\phi_0}=3.86$; $\chi=-0.5$, $p_{\phi_0}=4.20$ and $\chi=-0.9$, $p_{\phi_0}=4.36$ respectively.}\label{fig15}
\end{figure*}
The largest mass ratio of binary black holes investigated by numerical relativity is $100:1$ \cite{PhysRevLett.106.041101}. But the simulation only lasts two orbits which is too short to be used for gravitational waveform analysis. Till now the mass ratio of binary black hole investigated by numerical relativity for gravitational waveform usage is less than $20:1$ \cite{PhysRevD.93.044006,PhysRevD.93.044007,chu2016accuracy,PhysRevD.93.104050}. This limitation of numerical relativity is due to the computational cost for finite difference code, and/or due to the complicated computational grid adjustment for spectral code. In the future the finite element code may do some help on this problem \cite{PhysRevD.91.044033,miller2016operator,KIDDER201784}. But it is still under development. In contrast, our SEOBNRE model is free of this kind of limitation. This is similar to all other EOBNR models.

When the mass ratio becomes quite large, the small black hole can be looked as a perturbation source respect to the spacetime of the large black hole. Consequently the Teukolsky formalism is reasonable to treat this kind of binary problem. In \cite{han2011constructing} we constructed such a model to investigate extreme mass ratio binary system. In \cite{han2014gravitational}, one of us applied such model to investigate the eccentric binary black holes. It is interesting to compare the results got in \cite{han2014gravitational} and the ones simulated with the SEOBNRE model proposed here.

In all we have tested four cases. All of them are binary black holes with mass ratio about $1000:1$. More accurately the symmetric mass ratio is $\eta=10^{-3}$. The dynamical variables involved in the Teukolsky model \cite{han2011constructing,han2014gravitational} are the same to the ones in SEOBNRE model. So for the comparison in this subsection, we set the initial data for SEOBNRE model exactly the same to the ones for the Teukolsky model. More concretely, within spherical coordinate we set $r_0=\frac{p_0}{1+e_0}$ with $p_0=12$, $e_0=0.3$, $\phi_0=0$, $p_{r_0}=0$ and $p_{\phi_0}$ similar to the Fig.~4 of \cite{han2014gravitational}.

Not like the above test cases which involve slowly spinning black holes considered in previous subsections, the four test cases here admit high-spin black holes. We set the big black hole spinning while leave the small black hole spinless. The spin parameters $\chi$ for the four test cases are $0.9$, $0.5$, $-0.5$ and $-0.9$ respectively. Here the negative value means the spin direction is anti-paralleled to the direction of the orbital angular momentum of the binary. And the parameters $p_{\phi_0}$ for the initial data are $3.75295952324398$, $3.86330280736881$, $4.19862434393390$ and $4.35790829850906$ respectively. Different to the Fig.~4 of \cite{han2014gravitational}, here we have fixed $p_0=12$ while varied $p_{\phi_0}$ for corresponding $\chi$. This setting makes us easier to check the effect of $\chi$ on the gravitational waveform.

\begin{figure*}
\begin{tabular}{c}
\includegraphics[width=\textwidth]{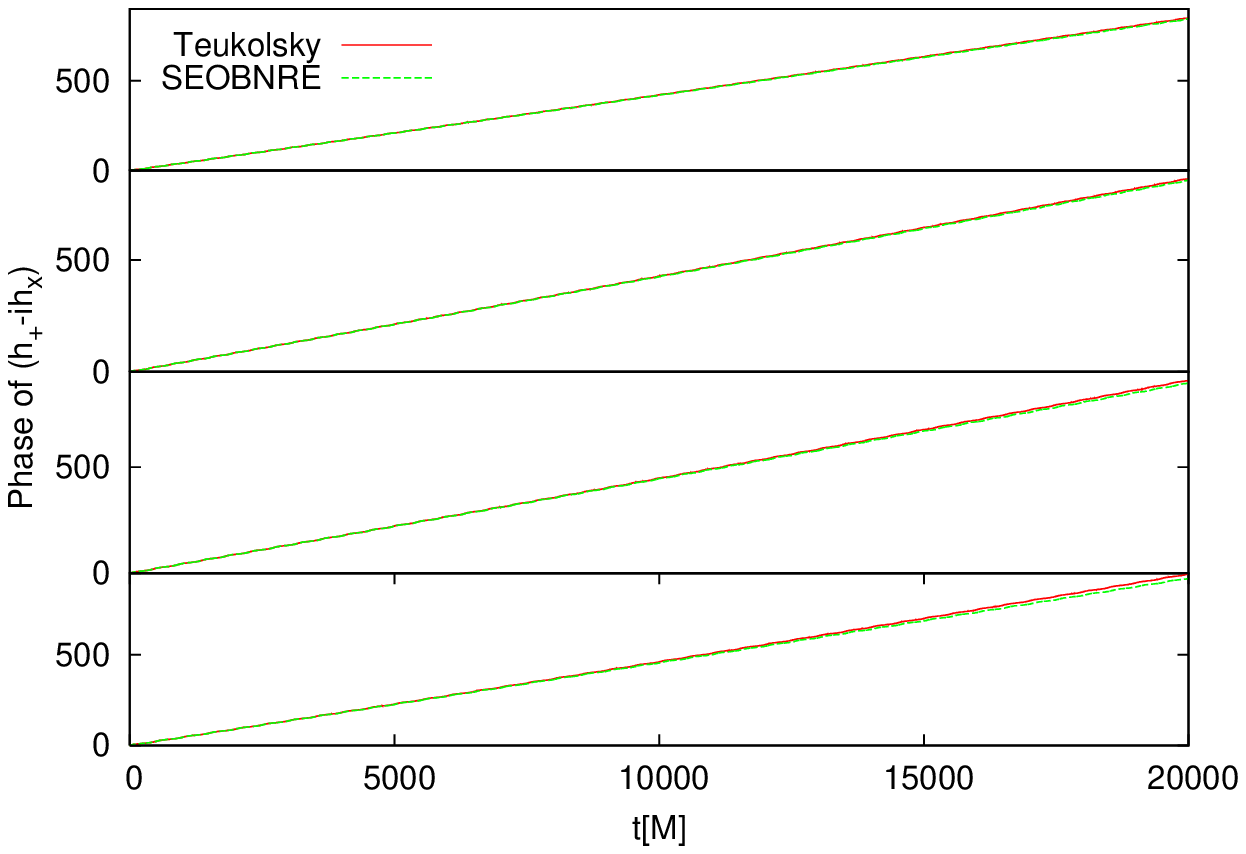}
\end{tabular}
\caption{Phase comparison of $h_{+}-ih_{\times}$ between Teukolsky equation based model and SEOBNRE model. Cases correspond to those of Fig.~\ref{fig15} respectively.}\label{fig16}
\end{figure*}

The gravitational waveform description adopted in \cite{han2014gravitational} used $h_{+,\times}$ (note the $y$-axis label typos involved in the Fig.~4 of \cite{han2014gravitational}). In order to make the comparison easier between the results in current paper and the ones in \cite{han2014gravitational}, we also adopt $h_{+,\times}$ to describe the gravitational waveform in this subsection. This is different to the spherical harmonic modes description used in previous subsections. Here we ignore the higher than 22 spherical harmonic modes and relate the $h_{+,\times}$ to $h_{22}$ through
\begin{align}
h_{+}-ih_{\times}=h_{22}{}^{-2}Y_{22}+h^{*}_{22}{}^{-2}Y_{2-2},
\end{align}
where we have used the relation $h_{2-2}=h^{*}_{22}$ with upper star denotes the complex conjugate \cite{PhysRevD.84.124052}. Then $h_{+,\times}$ are functions of direction angles. Following the Fig.~4 of \cite{han2014gravitational}, we plot $h_{+,\times}(\frac{\pi}{2},0)$ in Fig.~\ref{fig15}. Overall we can see that the consistence between the results of Teukolsky model and the ones of SEOBNRE model is good. When the comparison time becomes longer, the phase difference shows up. In Fig.~\ref{fig16} we compare the phase of $h_{+}-ih_{\times}$ corresponding to the cases shown in Fig.~\ref{fig15}.

As mentioned above, the four cases admit the same initial separation parameter $p_0=12$. Based on Newtonian gravity, the same mass ratio, the same initial separation and the same eccentric setting may reduce roughly the same orbital angular frequency. But we can see the larger $\chi$ ones admit a little bit faster frequency of gravitational waveform, while correspondingly the smaller orbital angular momentum $p_{\phi_0}$ ($3.75<3.86<4.20<4.36$). We attribute this to the faster frame dragging effect of the big black hole. In order to stay at the same radius position, corotating objects need smaller orbital angular momentum, but the anti-rotating objects need larger orbital angular momentum. And the frame dragging effect makes the gravitational wave frequency bigger for corotating case while smaller for anti-rotating case.

The overlap factors between the two respective waveforms for the cases shown in Fig.~\ref{fig15} are $\mathcal{O}=0.985367$, $0.985209$, $0.986240$, and $0.985558$ for $\chi=0.9$, $0.5$, $-0.5$ and $-0.9$ respectively. When the comparison time increases to 20000M, the overlap factors decrease to $0.690179$, $0.672447$, $0.549308$, and $0.498645$ respectively. One caution is in order here. Neither the SEOBNRE model nor the Teukolsky model is guaranteed to be accurate at 1 post-adiabatic order. So the good overlapping does not mean either model is accurate enough for gravitational wave detection usage. Since the two models make different
approximations and therefore introduce different errors, the good overlapping does imply those differences are ignorable for the compared cases. Only when the 1 post-adiabatic results are available, our SEOBNRE model can be checked more quantitatively for extreme mass ratio systems.
\section{Discussion and conclusion}\label{sec::discussion}
EOBNR model has contributed much to the gravitational wave detection. But the existing EOBNR models are limited to quasi-circular ($e=0$) systems. With no doubt, EOBNR model will continue to play an important role in the following LIGO observations. After about 20 years, space-based detectors will begin to work. It is interesting and important to ask whether the EOBNR model can still play an important role in the space-based gravitational wave detection. Among the gravitational wave sources for space-based detectors, binary systems are important. And many of such binary systems admit eccentric orbital motion \cite{0264-9381-24-17-R01}. Partial reason for this fact is that the decay rate of the eccentricity is proportional to the symmetric mass ratio. So although it may be not all of the issues limiting EOBNR model to work for space-based detectors, the eccentric orbit problem is an important point blocking EOBNR model to work for space-based detectors. So it is quite important to extend EOBNR model to describe eccentric binary systems. We proposed the first such extending model in the current paper--SEOBNRE model.

Our idea for constructing SEOBNRE model is combining the existing excellent property of EOBNR models for quasi-circular binary and the corrections coming from the eccentric orbit motion. The strategy is expanding involved quantities respect to eccentricity by assuming the smallness of the eccentricity. Then we calculate the correction terms coming from eccentricity through post-Newtonian approximation. Although the post-Newtonian order of the correction terms we got is only to second order, we expect that such kind of correction terms may work well. The reason is because when the eccentricity is large the separation of the binary is also large, then relative low order post-Newtonian approximation is necessary. Along with the decreasing of the separation of the binary, the eccentricity also decreases due to the circularizing effect of the gravitational radiation. Consequently the correction terms contribute weaker and weaker. In contrast, although the high PN order approximation is needed due to the decreasing of the separation, the existing excellent property of the EOBNR model can do the job.

Our SEOBNRE model includes Hamiltonian (Eq.~(\ref{SEOBNREHtotal})), waveform expression (Eqs.~(\ref{Ewave}) and (\ref{RDwaveform})) and the related energy flux (Eq.~(\ref{waveformflux})). We have compared our SEOBNRE model against to quasi-circular EOBNR model for consistency check. For $e=0$ the SEOBNRE model can recover the existing EOBNR models well. When the eccentricity $e$ increases, the difference between the SEOBNRE model and quasi-circular EOBNR model grows. We have introduced an overlap factor as defined in (\ref{overlapfactordef}) to quantify this difference. When $e<0.03$ the difference is small. The corresponding overlap factor is larger than 0.99. When $e>0.1$ the overlap factor becomes very low. As an example, the overlap factor for $e=0.3$ becomes 0.47. This result added evidences to the literature \cite{PhysRevD.92.044034} that quasi-circular template will break down when the eccentricity becomes larger than 0.1.

We have also compared SEOBNRE model against to another eccentric binary waveform model--ax model \cite{PhysRevD.95.024038}. When $e<0.15$ the overlap factor between the ax model and the SEOBNRE model is bigger than 0.9. This implies the consistence between the ax model and the SEOBNE model. As an example of $e>0.2$ cases, the overlap factor for $e=0.3$ between the ax model and the SEOBNRE model is as low as 0.67. This cautions us that more investigations are needed for waveform model about highly eccentric binary.

Numerical relativity (NR) simulation results can be looked as the standard answer for the eccentric binary. We have tested three numerical relativity simulation results. These three cases include spinless binary and spinning binary; equal mass binary and unequal mass binary. Compared to NR simulations with eccentricity 0.02, 0.02 and 0.19, the overlap factor for SEOBNRE model is 0.992, 0.997 and 0.982 respectively. In contrast, the overlap factor between NR waveform and $e=0$ SEOBNRE waveform is 0.990, 0.989 and 0.849 respectively.

Motivated by the gravitational wave sources for space-based detector, we have applied the SEOBNRE model to extreme mass ratio binaries. Specifically we considered binaries with mass ratio 1 to 1000. Since numerical relativity is not available yet for this kind of binaries, we compared the SEOBNRE model against to Teukolsky equation based model. Both spin aligned cases and anti-aligned cases are considered. All cases admit high eccentricity $e=0.3$. If we only care about time lasting several thousands $M$ the overlap factor between the SEOBNRE model and the Teukolsky equation based model is than 0.9. For total mass $M=10^6$ solar mass binaries several thousands $M$ corresponds to the time of hours. If we consider lasting time with tens of thousands $M$, the overlap factor drops below 0.7. Although the Teukolsky equation based model does not represent the standard answer as numerical relativity, this result also reminds us more work is needed when long duration time is involved.

In the current paper, simple overlap factor is considered to quantify the accuracy of the SEOBNRE model. For realistic gravitational wave detection, much more detail accuracy requirement \cite{PhysRevD.78.124020,PhysRevD.80.042005,PhysRevD.80.064019,PhysRevD.82.084020,PhysRevD.94.124030} is needed. It is interesting to ask if our SEOBNRE model is ready or not for particular detection projects such like eLISA \cite{amaro2012low}, LISA \cite{audley2017laser}, Taiji \cite{gong2011scientific} and Tianqin \cite{luo2016tianqin}. We leave such investigation as future works. On the other hand, there at least two possible clues to improve SEOBNRE model. The first one is calculating higher PN order terms for the eccentric correction. The second one is the trick adopted by Pan and his coworkers when they developed EOBNR model \cite{PhysRevD.76.104049}. The trick is adding some tuning parameters into the SEOBNRE model and then require these parameters to fit the calibration waveform such as numerical relativity simulation results. Noting that we did not put in any tunable parameters in the eccentric correction terms, there are two strategies to apply the mentioned trick. The first one is adjusting the existing parameters introduced in the original EOBNR models. The second one is adding more parameters into the eccentric correction terms and adjusting them.

Regarding to the merger and ringdown parts waveform (\ref{RDwaveform}), the possible improvement is taking the effect of the eccentricity on the mass and the spin of the final Kerr black hole into consideration. This point needs many more numerical relativity simulations to extend the relations (\ref{finalmass}) and (\ref{finalspin}) to include eccentricity. We leave these investigations to future study.
\acknowledgments
This work was supported by the NSFC (No.~11690023, No.~11622546, No.~11375260 and No.~U1431120). Z Cao was also supported by ``the Fundamental Research Funds for the Central Universities". W Han was also supported by Youth Innovation Promotion Association CAS and QYZDB-SSW-SYS016 of CAS.


\begin{widetext}
\appendix


\section{Detail expressions for the SEOBNRE Hamiltoanian}
\label{App:ham}
The SEOBNR Hamiltonian develops gradually, so the overall expressions are spreaded in the literature. In this appendix section, we collect all the results for SEOBNR Hamiltonian together. Our major references are \cite{PhysRevD.84.104027,PhysRevD.86.024011}. The involved terms corresponding to Eq.~(\ref{SEOBNREHterms}) can be expressed as following.
\begin{align}
\frac{H_{NS}}{M\eta}&=\beta^ip_i+\alpha\sqrt{1+\gamma^{ij}p_ip_j+Q_4(p)},\\
\frac{H_S}{M\eta}&=\omega S_{\hat{a}}
+\omega_r\frac{e^{-3\mu-\nu}\sqrt{\Delta_r}}{2B(1+\sqrt{Q})\sqrt{Q}\xi^2}[e^{2(\mu+\nu)}
p_\xi^2S_v r^2-Be^{\mu+\nu}p_vp_\xi S_\xi r^2\nonumber\\
&+B^2\xi^2(e^{2\mu}(\sqrt{Q}+Q)S_v
+p_np_vrS_n\sqrt{\Delta_r}-p_n^2S_v\Delta_r)]\nonumber\\
&+\omega_{\cos\theta}\frac{e^{-3\mu-\nu}\sqrt{\Delta_r}}{2B(1+\sqrt{Q})\sqrt{Q}}[S_n
(-e^{2(\mu+\nu)}p_\xi^2r^2+B^2(p_v^2r^2-e^{2\mu}(\sqrt{Q}+Q)\xi^2))-Bp_n(Bp_vS_v-e^{\mu+\nu}p_\xi S_\xi)r\sqrt{\Delta_r}]\nonumber\\
&+\frac{e^{-\mu+2\nu}}{B^2\sqrt{Q}\xi^2}(-B+e^{\mu+\nu})p_\xi S_{\hat{a}}r\nonumber\\
&+\frac{e^{-2\mu+\nu}}{B^2(\sqrt{Q}+Q)\xi^2}\{-Be^{\mu+\nu}\nu_{\cos\theta}p_\xi r(1+2\sqrt{Q})S_n\xi^2\nonumber\\
&+\sqrt{\Delta_r}[B(e^{\mu+\nu}\nu_rp_\xi r(1+2\sqrt{Q})S_v+B\mu_rp_vrS_\xi+BS_\xi(-\mu_{\cos\theta} p_n\xi^2+\sqrt{Q}(\mu_rp_vr-\nu_rp_vr+(\nu_{\cos\theta
}-\mu_{\cos\theta})p_n\xi^2)))\nonumber\\
&-B_re^{\mu+\nu}p_\xi r(1+\sqrt{Q})S_v]\},\\
H_{SC}&=-\frac{\eta}{2r^3}(S_*^2-3S_n^2)
+dh_{effSS}\frac{(M\eta)^2}{r^4}aS_{\hat{a}},\\
\alpha&=\frac{1}{\sqrt{-g^{tt}}},\beta^i=\frac{g^{ti}}{g^{tt}},\gamma^{ij}=g^{ij}-\frac{g^{ti}g^{tj}}{g^{tt}},
i,j=(r,\theta,\phi)\\
Q_4(p)&=2\eta(4-3\eta)\frac{\tilde{p}_n^4}{r^2},
Q=1+\frac{p_v^2r^2}{\Sigma\xi^2}
+\frac{\Delta_t\Sigma}{\Lambda_t}\frac{p_\xi^2r^2}{B^2\xi^2}
+p_n^2\frac{\Delta_r}{\Sigma},\\
S_*&=\sigma^*+\Delta^{(1)}_{\sigma_*}+\Delta^{(2)}_{\sigma_*}+\Delta^{(3)}_{\sigma_*},\sigma^*=a_1m_2+a_2m_1,\\
\Delta^{(1)}_{\sigma_*}&=\frac{\eta}{12r}\{\sigma[3(Q-1)r-8-36\bar{p}_n^2r]
+\sigma^*[4(Q-1)r+14-30\bar{p}_n^2r]\},\\
\Delta^{(2)}_{\sigma_*}&=\frac{\sigma}{144r^2}\{4(51\eta^2-109\eta)(Q-1)r
-16[7\eta(8+3\eta)]+810\eta^2\bar{p}_n^4r^2-45\eta(Q-1)^2r^2\nonumber\\
&-6\bar{p}_n^2r(16\eta+147\eta^2+(39\eta^2-6\eta)(Q-1)r)\}\nonumber\\
&-\frac{\sigma^*}{72r^2}\{2\eta(27\eta-353)+2(103\eta-60\eta^2)(Q-1)r
-360\eta^2\bar{p}_n^4r^2+(23+3\eta)\eta(Q-1)^2r^2\nonumber\\
&+6\bar{p}_n^2r[54\eta^2-47\eta+(21\eta^2-16\eta)(Q-1)r]\},\\
\Delta^{(3)}_{\sigma_*}&=\frac{d_{SO}\eta}{r^3}\sigma^*
\end{align}
where $S_*$ is the spin of the test particle deduced in the effective-one-body reduction. Following SEOBNRv1 we set $d_{SO}=-69.5, dh_{effSS}=2.75$. Our setting exactly follows the SEOBNRv1 code \cite{SEOBNRv1code}. In the above equations we have used notations
\begin{align}
\vec{n}&\equiv\frac{\vec{r}}{r},\vec{\xi}\equiv\frac{\vec{\sigma}}{\sigma}\times \vec{n},\vec{v}\equiv \vec{n}\times\vec{\xi},\label{xi26}\\
p_n&\equiv \vec{p}\cdot\vec{n}, p_\xi\equiv \vec{p}\cdot\vec{\xi}, p_v\equiv \vec{p}\cdot\vec{v},\\
S_n&\equiv \vec{S}_*\cdot\vec{n}, S_\xi\equiv \vec{S}_*\cdot\vec{\xi}, S_v\equiv \vec{S}_*\cdot\vec{v},\\
S_{\hat{a}}&\equiv \vec{S}_*\cdot\frac{\vec{a}}{a},
\bar{p}_n\equiv\sqrt{g^{rr}}p_n,\tilde{p}_n\equiv\vec{n}\cdot\vec{\tilde{p}}\\
\nu&=\frac{1}{2}\log(\frac{\Delta_t\Sigma}{\Lambda_t}),\mu=\frac{1}{2}\log(\Sigma),\\
\omega&\equiv\frac{\tilde{\omega}_{fd}}{\Lambda_t},B\equiv\sqrt{\Delta_t},\\
\nu_r&\equiv\frac{r}{\Sigma}+\frac{\bar{\omega}^2(\bar{\omega}^2\Delta_t'-4r\Delta_t)}{2\Lambda_t\Delta_t},
\mu_r\equiv\frac{r}{\Sigma}-\frac{1}{\sqrt{\Delta_r}},\\
\omega_r&\equiv\frac{\tilde{\omega}_{fd}'\Lambda_t-\tilde{\omega}_{fd}\Lambda_t'}{\Lambda_t^2},
B_r\equiv\frac{\sqrt{\Delta_r}\Delta_t'-2\Delta_t}{2\sqrt{\Delta_t\Delta_r}},\\
\nu_{\cos\theta}&\equiv\frac{a^2\bar{\omega}^2\cos\theta(\bar{\omega}^2-\Delta_t)}{\Lambda_t\Sigma},
\mu_{\cos\theta}\equiv\frac{a^2\cos\theta}{\Sigma},\\
\omega_{\cos\theta}&\equiv-\frac{2a^2\cos\theta\Delta_t\tilde{\omega}_{fd}}{\Lambda_t^2}
\end{align}
where the prime denotes the derivatives with respect to $r$. $(\vec{r},\vec{\tilde{p}})$ are the canonical variable within Boyer-Lindquist coordinate of the geodesic motion, while $\vec{p}$ is the momentum vector within tortoise coordinate. $\vec{p}$ and $\vec{\tilde{p}}$ are related through \cite{PhysRevD.81.084041}
\begin{align}
\vec{p}&=\vec{\tilde{p}}-\vec{n}(\vec{n}\cdot\vec{\tilde{p}})\frac{\xi_a-1}{\xi_a},\\
\xi_a&\equiv\frac{\sqrt{\Delta_t\Delta_r}}{r^2+a^2}.
\end{align}
Within spherical coordinate $\vec{r}=(r,\theta,\phi)$ and we set $\vec{a}$ along $\theta=0$ direction, so $\xi=\sin\theta$ in (\ref{xi26}).

\section{Detail expressions for the eccentric part of SEOBNRE waveform}
\label{App:waveform}
In this appendix section we show the detail calculation for the eccentric part of SEOBNRE waveform. We begin with the notations introduced in Eq.~(\ref{h22will}) as following.
\begin{align}
\Theta_{ij}&=\int(\epsilon_{ij}^{+}-i\epsilon_{ij}^{\times}){}^{-2}Y^*_{22}d\Omega,\\
P_n\Theta_{ij}&=\int N_n(\epsilon_{ij}^{+}-i\epsilon_{ij}^{\times}){}^{-2}Y^*_{22}d\Omega,\\
P_v\Theta_{ij}&=\int N_v(\epsilon_{ij}^{+}-i\epsilon_{ij}^{\times}){}^{-2}Y^*_{22}d\Omega,\\
P_{nn}\Theta_{ij}&=\int N_n^2(\epsilon_{ij}^{+}-i\epsilon_{ij}^{\times}){}^{-2}Y^*_{22}d\Omega,\\
P_{nv}\Theta_{ij}&=\int N_nN_v(\epsilon_{ij}^{+}-i\epsilon_{ij}^{\times}){}^{-2}Y^*_{22}d\Omega,\\
P_{vv}\Theta_{ij}&=\int N_v^2(\epsilon_{ij}^{+}-i\epsilon_{ij}^{\times}){}^{-2}Y^*_{22}d\Omega,\\
P_{nnn}\Theta_{ij}&=\int N_n^3(\epsilon_{ij}^{+}-i\epsilon_{ij}^{\times}){}^{-2}Y^*_{22}d\Omega,\\
P_{nnv}\Theta_{ij}&=\int N_n^2N_v(\epsilon_{ij}^{+}-i\epsilon_{ij}^{\times}){}^{-2}Y^*_{22}d\Omega,\\
P_{nvv}\Theta_{ij}&=\int N_nN_v^2(\epsilon_{ij}^{+}-i\epsilon_{ij}^{\times}){}^{-2}Y^*_{22}d\Omega,\\
P_{vvv}\Theta_{ij}&=\int N_v^3(\epsilon_{ij}^{+}-i\epsilon_{ij}^{\times}){}^{-2}Y^*_{22}d\Omega.
\end{align}
Based on Eqs.~(\ref{ewaveb})-(\ref{ewavee}), direct calculation gives
\begin{align}
\Theta_{ij}&=(\frac{1}{3}\sqrt{\frac{\pi }{5}},-\frac{i}{3}\sqrt{\frac{\pi }{5}},0,-\frac{1}{3}\sqrt{\frac{\pi }{5}},\frac{8}{3 \sqrt{5 \pi }},0),\\
P_n\Theta_{ij}&=(0,0,-\frac{\sqrt{\frac{\pi }{5}} (x_1-i x_2)}{3 \sqrt{x_1^2+x_2^2+x_3^2}},0,\frac{\sqrt{\frac{\pi }{5}} (i x_1+x_2)}{3 \sqrt{x_1^2+x_2^2+x_3^2}},0),\\
P_v\Theta_{ij}&=(0,0,-\frac{1}{3} \sqrt{\frac{\pi }{5}} (v_1-i v_2),
0,\frac{1}{3} \sqrt{\frac{\pi }{5}} (i v_1+v_2),0),\\
P_{nn}\Theta_{ij}&=(\frac{\sqrt{\frac{\pi }{5}} \left(-x_1^2+8 i x_1 x_2+7 x_2^2+x_3^2\right)}{21 \left(x_1^2+x_2^2+x_3^2\right)},
-\frac{i \sqrt{\frac{\pi }{5}} \left(3 x_1^2+3 x_2^2+x_3^2\right)}{21 \left(x_1^2+x_2^2+x_3^2\right)},\frac{2 \sqrt{\frac{\pi }{5}} (x_1-i x_2) x_3}{21 \left(x_1^2+x_2^2+x_3^2\right)},\nonumber\\
&\frac{\sqrt{\frac{\pi }{5}} \left(-7 x_1^2+8 i x_1 x_2+x_2^2-x_3^2\right)}{21 \left(x_1^2+x_2^2+x_3^2\right)},-\frac{2 i \sqrt{\frac{\pi }{5}} (x_1-i x_2) x_3}{21 \left(x_1^2+x_2^2+x_3^2\right)},\frac{8 \sqrt{\frac{\pi }{5}} (x_1-i x_2)^2}{21 \left(x_1^2+x_2^2+x_3^2\right)}),\\
P_{nv}\Theta_{ij}&=(\frac{\sqrt{\frac{\pi }{5}} (-v_1 x_1+4 i v_2 x_1+4 i v_1 x_2+7 v_2 x_2+v_3 x_3)}{21 \sqrt{x_1^2+x_2^2+x_3^2}},-\frac{i \sqrt{\frac{\pi }{5}} (3 v_1 x_1+3 v_2 x_2+v_3 x_3)}{21 \sqrt{x_1^2+x_2^2+x_3^2}},\nonumber\\
&\frac{\sqrt{\frac{\pi }{5}} (v_3 (x_1-i x_2)+(v_1-i v_2) x_3)}{21 \sqrt{x_1^2+x_2^2+x_3^2}},\frac{\sqrt{\frac{\pi }{5}} (-7 v_1 x_1+4 i v_2 x_1+4 i v_1 x_2+v_2 x_2-v_3 x_3)}{21 \sqrt{x_1^2+x_2^2+x_3^2}},\nonumber\\
&-\frac{\sqrt{\frac{\pi }{5}} (v_3 (i x_1+x_2)+(i v_1+v_2) x_3)}{21 \sqrt{x_1^2+x_2^2+x_3^2}},\frac{8 \sqrt{\frac{\pi }{5}} (v_1-i v_2) (x_1-i x_2)}{21 \sqrt{x_1^2+x_2^2+x_3^2}}),\\
P_{vv}\Theta_{ij}&=(\frac{1}{21} \sqrt{\frac{\pi }{5}} \left(-v_1^2+8 i v_1 v_2+7 v_2^2+v_3^2\right),-\frac{1}{21} i \sqrt{\frac{\pi }{5}} \left(3 v_1^2+3 v_2^2+v_3^2\right),\nonumber\\
&\frac{2}{21} \sqrt{\frac{\pi }{5}} (v_1-i v_2) v_3,\frac{1}{21} \sqrt{\frac{\pi }{5}} \left(-7 v_1^2+8 i v_1 v_2+v_2^2-v_3^2\right),\nonumber\\
&-\frac{2}{21} i \sqrt{\frac{\pi }{5}} (v_1-i v_2) v_3,\frac{8}{21} \sqrt{\frac{\pi }{5}} (v_1-i v_2)^2),\\
P_{nnn}\Theta_{ij}&=(\frac{\sqrt{\frac{\pi }{5}} (x_1-i x_2)^2 x_3}{7 \left(x_1^2+x_2^2+x_3^2\right)^{3/2}},0,\nonumber\\
&-\frac{\sqrt{\frac{\pi }{5}} (x_1-i x_2) \left(2 x_1^2+5 i x_1 x_2+7 x_2^2+3 x_3^2\right)}{21 \left(x_1^2+x_2^2+x_3^2\right)^{3/2}},\frac{\sqrt{\frac{\pi }{5}} (x_1-i x_2)^2 x_3}{7 \left(x_1^2+x_2^2+x_3^2\right)^{3/2}},\nonumber\\
&\frac{\sqrt{\frac{\pi }{5}} (i x_1+x_2) \left(7 x_1^2-5 i x_1 x_2+2 x_2^2+3 x_3^2\right)}{21 \left(x_1^2+x_2^2+x_3^2\right)^{3/2}},-\frac{2 \sqrt{\frac{\pi }{5}} (x_1-i x_2)^2 x_3}{7 \left(x_1^2+x_2^2+x_3^2\right)^{3/2}}),\\
P_{nnv}\Theta_{ij}&=(\frac{\sqrt{\frac{\pi }{5}} (x_1-i x_2) (v_3 (x_1-i x_2)+2 (v_1-i v_2) x_3)}{21 \left(x_1^2+x_2^2+x_3^2\right)},0,\nonumber\\
&-\frac{\sqrt{\frac{\pi }{5}} \left((x_1-i x_2) (2 v_1 x_1+i v_2 x_1+4 i v_1 x_2+7 v_2 x_2)+2 v_3 (x_1-i x_2) x_3+(v_1-i v_2) x_3^2\right)}{21 \left(x_1^2+x_2^2+x_3^2\right)},\nonumber\\
&\frac{\sqrt{\frac{\pi }{5}} (x_1-i x_2) (v_3 (x_1-i x_2)+2 (v_1-i v_2) x_3)}{21 \left(x_1^2+x_2^2+x_3^2\right)},\nonumber\\
&\frac{\sqrt{\frac{\pi }{5}} \left((x_1-i x_2) (7 i v_1 x_1+4 v_2 x_1+v_1 x_2+2 i v_2 x_2)+2 v_3 (i x_1+x_2) x_3+(i v_1+v_2) x_3^2\right)}{21 \left(x_1^2+x_2^2+x_3^2\right)},\nonumber\\
&-\frac{2 \sqrt{\frac{\pi }{5}} (x_1-i x_2) (v_3 (x_1-i x_2)+2 (v_1-i v_2) x_3)}{21 \left(x_1^2+x_2^2+x_3^2\right)}),\\
P_{nvv}\Theta_{ij}&=(\frac{\sqrt{\frac{\pi }{5}} (v_1-i v_2) (2 v_3 (x_1-i x_2)+(v_1-i v_2) x_3)}{21 \sqrt{x_1^2+x_2^2+x_3^2}},0,\nonumber\\
&-\frac{\sqrt{\frac{\pi }{5}} \left(v_3^2 (x_1-i x_2)+v_1^2 (2 x_1+i x_2)+v_2^2 (4 x_1-7 i x_2)-2 i v_2 v_3 x_3+2 v_1 (i v_2 x_1+4 v_2 x_2+v_3 x_3)\right)}{21 \sqrt{x_1^2+x_2^2+x_3^2}},\nonumber\\
&\frac{\sqrt{\frac{\pi }{5}} (v_1-i v_2) (2 v_3 (x_1-i x_2)+(v_1-i v_2) x_3)}{21 \sqrt{x_1^2+x_2^2+x_3^2}},\nonumber\\
&\frac{\sqrt{\frac{\pi }{5}} \left(v_3^2 (i x_1+x_2)+v_2^2 (-i x_1+2 x_2)+v_1^2 (7 i x_1+4 x_2)+2 v_2 v_3 x_3+v_1 (8 v_2 x_1-2 i v_2 x_2+2 i v_3 x_3)\right)}{21 \sqrt{x_1^2+x_2^2+x_3^2}},\nonumber\\
&-\frac{2 \sqrt{\frac{\pi }{5}} (v_1-i v_2) (2 v_3 (x_1-i x_2)+(v_1-i v_2) x_3)}{21 \sqrt{x_1^2+x_2^2+x_3^2}}),\\
P_{vvv}\Theta_{ij}&=(\frac{1}{7} \sqrt{\frac{\pi }{5}} (v_1-i v_2)^2 v_3,0,\nonumber\\
&-\frac{1}{21} \sqrt{\frac{\pi }{5}} (v_1-i v_2) \left(2 v_1^2+5 i v_1 v_2+7 v_2^2+3 v_3^2\right),\frac{1}{7} \sqrt{\frac{\pi }{5}} (v_1-i v_2)^2 v_3,\nonumber\\
&\frac{1}{21} \sqrt{\frac{\pi }{5}} (i v_1+v_2) \left(7 v_1^2-5 i v_1 v_2+2 v_2^2+3 v_3^2\right),-\frac{2}{7} \sqrt{\frac{\pi }{5}} (v_1-i v_2)^2 v_3)
\end{align}
In the above equations, we have listed by the components order $11$, $12$, $13$, $22$, $23$, and $33$ within Cartesian coordinate. The position and velocity components mean $\vec{r}=(x_1,x_2,x_3), \vec{v}_p=(v_1,v_2,v_3)$. And more we have
\begin{align}
P_n^{\frac{1}{2}}Q^{ij}&=\frac{3(m_1-m_2)}{r}(n^iv_p^j+v_p^in^j-\dot{r}n^in^j),\\
P_v^{\frac{1}{2}}Q^{ij}&=(m_1-m_2)(\frac{n^in^j}{r}-2v_p^iv_p^j),\\
P_0Q^{ij}&=\frac{1}{3}\{[3(1-3\eta)v_p^2-2\frac{(2-3\eta)}{r}]v_p^iv_p^j
+\frac{2}{r}\dot{r}(5+3\eta)(n^iv_p^j+v_p^in^j)+[3(1-3\eta)\dot{r}^2
-(10+3\eta)v_p^2+\frac{29}{r}]\frac{n^in^j}{r}\},\\
P_{nn}Q^{ij}&=\frac{1-3\eta}{3r}[(3v_p^2-15\dot{r}^2+\frac{7}{r})n^in^j
+15\dot{r}(n^iv_p^j+v_p^in^j)-14v_p^iv_p^j],\\
P_{nv}Q^{ij}&=\frac{1-3\eta}{3r}[12\dot{r}n^in^j-16(n^iv_p^j+v_p^in^j)],\\
P_{vv}Q^{ij}&=\frac{1-3\eta}{3}(6v_p^iv_p^j-\frac{2}{r}n^in^j),\\
P_n^{\frac{3}{2}}Q^{ij}&=\frac{(m_1-m_2)}{12r}\{(n^iv_p^j+v_p^in^j)[\dot{r}^2(63+54\eta)-\frac{128-36\eta}{r}
+v_p^2(33-18\eta)]\nonumber\\
&+n^in^j\dot{r}[\dot{r}^2(15-90\eta)-v_p^2(63-54\eta)+\frac{242-24\eta}{r}]
-\dot{r}v_p^iv_p^j(186+24\eta)\},\\
P_v^{\frac{3}{2}}Q^{ij}&=(m_1-m_2)\{\frac{1}{2}v_p^iv_p^j[\frac{3-8\eta}{r}-2v_p^2(1-5\eta)]
-\frac{n^iv^j+v_p^in^j}{2r}\dot{r}(7+4\eta)\nonumber\\
&-\frac{n^in^j}{r}[\frac{3}{4}(1-2\eta)\dot{r}^2
+\frac{1}{3}\frac{26-3\eta}{r}-\frac{1}{4}(7-2\eta)v_p^2]\},\\
P_{nnn}^{\frac{3}{2}}Q^{ij}&=\frac{(m_1-m_2)(1-2\eta)}{r}[\frac{5}{4}(3v_p^2-7\dot{r}^2+\frac{6}{r})\dot{r}n^in^j
-\frac{17}{2}\dot{r}v_p^iv_p^j-(21v_p^2-105\dot{r}^2
+\frac{44}{r})\frac{n^iv_p^j+v_p^in^j}{12}],\\
P_{nnv}^{\frac{3}{2}}Q^{ij}&=\frac{(m_1-m_2)(1-2\eta)}{4r}[58v_p^iv_p^j+(45\dot{r}^2-9v_p^2-\frac{28}{r})n^in^j
-54\dot{r}(n^iv_p^j+v_p^in^j)],\\
P_{nvv}^{\frac{3}{2}}Q^{ij}&=\frac{3(m_1-m_2)(1-2\eta)}{2r}(5(n^iv_p^j+v_p^in^j)-3\dot{r}n^in^j),\\
P_{vvv}^{\frac{3}{2}}Q^{ij}&=\frac{(m_1-m_2)(1-2\eta)}{2}(\frac{n^in^j}{r}-4v_p^iv_p^j).
\end{align}

\section{Post-Newtonian energy flux for eccentric binary}
\label{App:PNenergy}
In this appendix section, we show the Post-Newtonian energy flux for eccentric binary. Following \cite{PhysRevD.95.024038} for the circular part we can combine the results involved in SEOBNRv1 \cite{Cao16,PhysRevD.85.064010,PhysRevD.95.024038} to get
\begin{align}
-\frac{dE}{dt}|_{(C)}=&\frac{32}{5}\eta^2 x^5[1-\frac{1247+980\eta}{336}x+4\pi x^{3/2}+(-\frac{44711}{9072}+\frac{9271\eta}{504}+\frac{65\eta^2}{18})x^2
-(\frac{8191}{672}+\frac{583\eta}{24})\pi x^{5/2}\nonumber\\
&+(\frac{6643739519}{69854400}-\frac{1712\gamma_E}{105}+\frac{16 \pi ^2}{3}-\frac{134543 \eta }{7776}+\frac{41 \pi ^2 \eta }{48}-\frac{94403 \eta ^2}{3024}-\frac{775 \eta ^3}{324}-\frac{856 \log(16)}{105}-\frac{856 \log(x)}{105})x^3\nonumber\\
&+(-\frac{16285}{504}+\frac{214745\eta }{1728}+\frac{27755\eta ^2}{432})\pi x^{7/2}\nonumber\\
&+(-\frac{23971119313}{93139200}+\frac{856 \gamma_E}{35}-8 \pi ^2-\frac{59292668653 \eta }{838252800}+5 a_0 \eta +\frac{856 \gamma \eta }{315}+\frac{31495 \pi ^2 \eta }{8064}-\frac{54732199 \eta ^2}{93312}+\frac{3157 \pi ^2 \eta ^2}{144}\nonumber\\
&+\frac{18929389 \eta ^3}{435456}+\frac{97 \eta ^4}{3888}+\frac{428 \log(16)}{35}+\frac{428}{315} \eta  \log(16)+\frac{428 \log(x)}{35}+\frac{47468}{315} \eta  \log(x))x^4\nonumber\\
&+(-\frac{80213 }{768}+\frac{51438847  \eta }{48384}-\frac{205 \pi ^2 \eta }{6}-\frac{42745411  \eta ^2}{145152}-\frac{4199 \eta ^3}{576})\pi x^{9/2}\nonumber\\
&+(-\frac{121423329103}{82790400}+\frac{5778 \gamma_E}{35}-54 \pi ^2+\frac{4820443583363 \eta }{1257379200}-\frac{3715 a_0 \eta }{336}+6 a_1 \eta -\frac{4066 \gamma_E \eta }{35}-\frac{31869941 \pi ^2 \eta }{435456}\nonumber\\
&-\frac{2006716046219 \eta ^2}{3353011200}-\frac{55 a_0 \eta ^2}{4}+\frac{214 \gamma_E \eta ^2}{105}+\frac{406321 \pi ^2 \eta ^2}{48384}+\frac{2683003625 \eta ^3}{3359232}-\frac{100819 \pi ^2 \eta ^3}{3456}-\frac{192478799 \eta ^4}{5225472}\nonumber\\
&+\frac{33925 \eta ^5}{186624}+\frac{2889 \log(16)}{35}-\frac{2033}{35} \eta  \log(16)+\frac{107}{105} \eta ^2 \log(16)+\frac{2889 \log(x)}{35}-\frac{391669}{315} \eta  \log(x)\nonumber\\
&-\frac{122981}{105} \eta ^2 \log(x))x^5\nonumber\\
&+(-\frac{623565 }{1792}-\frac{235274549 \eta }{241920}+20 a_0  \eta +\frac{852595 \pi ^2 \eta }{16128}-\frac{187219705 \eta ^2}{32256}+\frac{12915 \pi ^2 \eta ^2}{64}+\frac{503913815 \eta ^3}{870912}\nonumber\\
&-\frac{24065 \eta ^4}{3456}+\frac{1792}{3} \eta  \log(x)) \pi x^{11/2}\nonumber\\
&+(-\frac{1216355221}{206976}+\frac{4815 \gamma_E}{7}-225 \pi ^2+\frac{45811843687349 \eta }{1149603840}+\frac{170515 a_0 \eta }{18144}-\frac{743 a_1 \eta }{56}+7 a_2 \eta +a_3 \eta \nonumber\\
&-\frac{737123 \gamma_E \eta }{189}-\frac{84643435883 \pi ^2 \eta }{670602240}+\frac{8774}{63} \gamma_E \pi ^2 \eta -\frac{410 \pi ^4 \eta }{9}-\frac{37516325949517 \eta ^2}{603542016}+\frac{68305 a_0 \eta ^2}{2016}-\frac{33 a_1 \eta ^2}{2}\nonumber\\
&+\frac{6634 \gamma_E \eta ^2}{63}+\frac{23084972185 \pi ^2 \eta ^2}{5225472}-\frac{92455 \pi ^4 \eta ^2}{1152}+\frac{6069288163291 \eta ^3}{2586608640}+\frac{295 a_0 \eta ^3}{18}+\frac{107 \gamma_E \eta ^3}{243}-\frac{114930545 \pi ^2 \eta ^3}{1741824}\nonumber\\
&-\frac{145089945295 \eta ^4}{282175488}+\frac{141655 \pi ^2 \eta ^4}{7776}+\frac{6942085 \eta ^5}{497664}+\frac{196175 \eta ^6}{3359232}+\frac{4815 \log(16)}{14}-\frac{737123}{378} \eta  \log(16)\nonumber\\
&+\frac{4387}{63} \pi ^2 \eta  \log(16)+\frac{3317}{63} \eta ^2 \log(16)+\frac{107}{486} \eta ^3 \log(16)+\frac{4815 \log(x)}{14}+\frac{13185899 \eta  \log(x)}{59535}+7 a_3 \eta  \log(x)\nonumber\\
&+\frac{4387}{63} \pi ^2 \eta  \log(x)+\frac{963937}{189} \eta ^2 \log(x)+\frac{6279367 \eta ^3 \log(x)}{2430})x^6\nonumber\\
&+(-2\chi_S-\frac{3}{4}\sqrt{1-4\eta}\chi_A)x^{3/2}\nonumber\\
&+[(-\frac{9}{4}+\frac{136}{9}\eta)\chi_S+(-\frac{23}{16}
+\frac{157}{36}\eta)\sqrt{1-4\eta}\chi_A]x^{5/2}\nonumber\\
&+(-8\pi\chi_S-\frac{17}{6}\pi\sqrt{1-4\eta}\chi_A)x^3\nonumber\\
&+((\frac{476645}{13608}+\frac{3086}{189}\eta-\frac{1405}{27}\eta^2)\chi_S
+(\frac{180955}{27216}+\frac{625}{378}\eta-\frac{1117}{108}\eta^2)\sqrt{1-4\eta}\chi_A)x^{7/2}],\label{energyfluxwo}
\end{align}
with $a_0=153.8803,a_2=-55.13,a_2=588,a_3=-1144$. These $a$ values are taken from \cite{PhysRevD.85.064010,PhysRevD.95.024038}. Here $x\equiv\sqrt{v}$. The PN results for elliptic orbit read \cite{PhysRevD.52.821,PhysRevD.54.4813}
\begin{align}
-\frac{dE}{dt}|_{Elip}=&\frac{8}{15}\frac{M^4\eta^2}{r^4}[12v^2-11\dot{r}^2\nonumber\\
&+\frac{1}{28} (-2 \dot{r}^2 v^2 (1487-1392 \eta )+v^4 (785-852 \eta )+3 \dot{r}^4 (687-620 \eta )+\frac{8 M \dot{r}^2 (367-15 \eta )}{r}\nonumber\\
&+\frac{16 M^2 (1-4 \eta )}{r^2}-\frac{160 M v^2 (17-\eta )}{r})\nonumber\\
&+\frac{1}{756} (-\frac{24 M^3 (253-1026 \eta +56 \eta ^2)}{r^3}-\frac{36 M v^4 (4446-5237 \eta +1393 \eta ^2)}{r}\nonumber\\
&+\frac{108 M \dot{r}^2 v^2 (4987-8513 \eta +2165 \eta ^2)}{r}+\frac{M^2 v^2 (281473+81828 \eta +4368 \eta ^2)}{r^2}+18 v^6 (1692-5497 \eta +4430 \eta ^2)\nonumber\\
&-\frac{3 M^2 \dot{r}^2 (106319+9798 \eta +5376 \eta ^2)}{r^2}-54 \dot{r}^2 v^4 (1719-10278 \eta +6292 \eta ^2)+54 \dot{r}^4 v^2 (2018-15207 \eta +7572 \eta ^2)\nonumber\\
&-18 \dot{r}^6 (2501-20234 \eta +8404 \eta ^2)-\frac{12 M \dot{r}^4 (33510-60971 \eta +14290 \eta ^2)}{r})]\nonumber\\
&+\frac{8}{15}\frac{M^5\eta^2}{r^5}(\hat{n}\times\vec{v})\cdot[(\vec{\chi}_S
+\sqrt{1-4\eta}\vec{\chi}_A)(27\dot{r}^2-37v^2-12\frac{M}{r})
+4\eta\vec{\chi}_S(12\dot{r}^2-3v^2+8\frac{M}{r})]\nonumber\\
&+\frac{8}{15}\frac{M^6\eta^3}{r^6}[3(\chi_S^2-\chi_A^2)(47v^2-55\dot{r}^2)
-3((\hat{n}\cdot\vec{\chi}_S)^2+(\hat{n}\cdot\vec{\chi}_A)^2)(168v^2-269\dot{r}^2)
+71((\vec{v}\cdot\vec{\chi}_S)^2+(\vec{v}\cdot\vec{\chi}_A)^2)\nonumber\\
&-342\dot{r}[(\vec{v}\cdot\vec{\chi}_S)(\hat{n}\cdot\vec{\chi}_S)
-(\vec{v}\cdot\vec{\chi}_A)(\hat{n}\cdot\vec{\chi}_A)]],
\end{align}
where the PN order for the non-spin part and spin-spin interaction part is second, but the PN order for the spin-orbit interaction part is only 1.5.
\end{widetext}
\bibliography{refs}

\end{document}